# Intermetallic Nanocrystals: Syntheses and Catalytic Applications

*Yucong Yan, Jingshan S. Du, Kyle D. Gilroy, Deren Yang, Younan Xia,\* and Hui Zhang\**

Y. Yan, J. S. Du,[+] Prof. D. Yang, Prof. H. Zhang

State Key Laboratory of Silicon Materials, School of Materials Science and Engineering, Zhejiang University, Hangzhou, Zhejiang 310027, P. R. China

E-mail: msezhanghui@zju.edu.cn

[+] Present address: Department of Materials Science and Engineering, Northwestern University, Evanston, Illinois 60201, United States

Dr. K. D. Gilroy, Prof. Y. Xia

The Wallace H. Coulter Department of Biomedical Engineering, Georgia Institute of Technology and Emory University, Atlanta, Georgia 30332, United States

Email: younan.xia@bme.gatech.edu

Prof. Y. Xia

School of Chemistry and Biochemistry, School of Chemical and Biomolecular Engineering, Georgia Institute of Technology, Atlanta, Georgia 30332, United States

## Abstract

At the forefront of nanochemistry, there exists a research endeavor centered around intermetallic nanocrystals, which are unique in terms of long-range atomic ordering, well-defined stoichiometry, and controlled crystal structure. In contrast to alloy nanocrystals with no atomic ordering, it has been challenging to synthesize intermetallic nanocrystals with a tight control over their size and shape. This review article highlights recent progress in the synthesis of intermetallic nanocrystals with controllable sizes and well-defined shapes. We begin with a simple analysis and some insights key to the selection of experimental conditions for generating intermetallic nanocrystals. We then present examples to highlight the viable use of intermetallic nanocrystals as electrocatalysts or catalysts for various reactions, with a focus on the enhanced performance relative to their alloy counterparts that lack atomic ordering. We conclude with perspectives on future developments in the context of synthetic control, structure-property relationship, and application.

# 1. Introduction

Metals, covering more than 75% of the elements in the periodic table, have played a central role in modernizing human civilization due to their unique physicochemical properties. This is evident from their widespread use as structural materials for buildings, bridges, airplanes, and vehicles; as conductors for electrical systems and electronic devices; and as biomaterials for medical implants, diagnostic probes/tools, and therapeutic agents. About 160 years ago, it was discovered by Michael Faraday that the scope and application space of metals could be further expanded when processed in the form of nanomaterials. In a finely divided state, as in the form of nanocrystals, metals exhibit intriguing electronic, optical, magnetic, and catalytic properties different from the bulk materials.[1-17] Over the past two decades, synthetic methods have also been developed to control both the size and shape of metal nanocrystals.[18-31] Access to such well-defined nanomaterials has enabled structure-property studies, including the identification of optimal surface compositions and structures for specific catalytic reactions.[32-39] One such example was reported by Somorjai and coworkers, who demonstrated that Pt cuboctahedral nanocrystals, enclosed by a mix of {111} and {100} facets, produced both cyclohexene and cyclohexane during benzene hydrogenation, whereas only cyclohexane was generated when Pt cubic nanocrystals, encased by {100} facets, were used.[39] This remarkable discovery, along with many others, stimulated tremendous efforts in the development of shape-controlled metal nanocrystals for catalytic applications.[40-44]

Despite the remarkable success, nanocrystals comprised of a single metal simply cannot meet all the requirements for a given catalytic application. Ideally, a catalyst should have high activity and selectivity, in addition to chemical and structural stability, while being available at low cost.[45-52] All these essential criteria can be potentially met by exploiting the synergy that will arise when a second metal is incorporated for the formation of bimetallic nanocrystals. By introducing a second metal, the property landscape will be drastically expanded, allowing for unprecedented improvement in terms of catalytic performance.[53-62] One notable example can be found in the strategy that has been widely explored for enhancing the activity of Pt toward the oxygen reduction reaction (ORR) by mixing it with a 3*d* transition metal such as Ni, Co, or Cu.[63-67] In general, the superior performance can be attributed to one or more of the effects that arise when two metals are involved, including the *ensemble effect* -- where specific groups of surface atoms take on distinct mechanistic functionalities; the *ligand effect* -- where charge transfer between two dissimilar surface atoms alters their electronic structure and activity; and the *geometric effect* -- where the spatial arrangement of surface atoms is affected by strain, geometry, and size.[68-73] To this end, many research groups have demonstrated the ability to maneuver the size, shape (or the types of facets), architecture (core-shell *vs.* nanoframe), in addition to composition (elemental makeup and atomic ratio) and crystal structure (arrangement of atoms or unit cell type).[73] Any variation to the first three parameters can result in markedly different catalytic properties even if the nanocrystals share exactly the same chemical composition or atomic ratio.[74-76]

It is simply inadequate to use the term "composition" to describe bimetallic nanocrystals as their properties are highly sensitive to the nature of atomic ordering (*i.e.*, random alloy *vs.* intermetallic compound) even when they are exactly identical in terms of composition and stoichiometry.[77-79] This difference can be attributed to the changes in crystal structure and surface structure, which can modify the strength and nature of how chemical species adsorb and react, respectively, leading to different catalytic properties.[80-84] In a system involving Pt and Fe, extended X-ray absorption fine structure (EXAFS) analysis indicated that both Pt-Pt and Pt-Fe bond lengths were shortened when converting disordered face-centered cubic (*fcc*, A1; space group: $Fm\bar{3}m$) PtFe nanocrystals into ordered, face-centered tetragonal (*fct*, L1$_0$; space group: $P4/mmm$) PtFe nanocrystals, resulting in a lower oxygen adsorption energy and thus enhanced ORR activity.[80] Intermetallics also greatly benefit from the ordered nature of the atoms, as demonstrated by DiSalvo and coworkers, who compared the catalytic activities of Pt$_3$Ti intermetallic nanoparticles with those made of a random alloy for the oxidation of methanol and formic acid, with the intermetallic phase showing superior performance due to much lower affinity for CO adsorption.[77] In some cases, the enhancement in catalytic performance is closely related to the electronic modification and isolation of active sites that can substantially enhance the selectivity of a catalytic reaction (*i.e*., site-isolation effect), a concept that has been widely used in the semi-hydrogenation of alkyne groups.[82,83] In addition to the superior activity and selectivity is the remarkable stability of intermetallics because they generally perform better in harsh chemical environments, during electrochemical cycling, and upon exposure to high temperatures. This enhanced stability can be attributed to the heteroatomic bonding in intermetallics, which tends to have a more negative enthalpy of formation relative to the random alloy. For Pt-based intermetallic catalysts containing non-noble metals (*e.g.*, Al, Fe, and Cu),[85-87] recent studies indicate that the ordered phases showed significantly less leaching of their non-noble metal components than their alloy counterparts with identical composition when electrochemically cycled in acidic solutions.

Inspired by their remarkable activity, selectivity, and durability toward a broad range of catalytic reactions, intermetallic nanocrystals have been a subject of active research in recent years.[88-92] While thermal annealing of a dry sample is often used to produce intermetallic nanomaterials,[93-95] it leaves no room for shape control. In contrast, solution-phase synthesis stands out as the foremost route to the generation of intermetallic nanocrystals with well-controlled sizes, shapes, and internal defect structures.[96-100] In this *Review*, we categorize synthetic strategies into thermal annealing in a gas phase (or vacuum) and wet-chemical synthesis in a solution. Specifically, we cover techniques capable of generating intermetallic structures through *i*) thermal annealing of alloys or heterostructures in a reducing or inert atmosphere (or vacuum) at temperatures lower than the order-to-disorder transition point; *ii*) wet-chemistry approaches including seed-mediated diffusion growth in the presence of preformed seeds or direct one-pot syntheses involving the simultaneous reduction and/or decomposition of two metal precursors. We place a special emphasis on the key experimental parameters and mechanisms that govern the formation of intermetallic nanocrystals. After highlighting their widespread use as electrocatalysts or catalysts for a variety of reactions, we conclude by discussing some of the potential avenues one could take to address the current challenges associated with controlling the synthesis of intermetallic nanocrystals.

## 2. Thermodynamic and kinetic perspective on intermetallics

Bimetallic nanocrystals produced through wet-chemical syntheses typically take a random, alloy structure and are thus more prevalent than the intermetallic variants, even when the latter is favored thermodynamically. This outcome can be attributed to the relatively low temperatures intrinsic to most wet-chemical syntheses compared to the high activation energy barrier to inter-diffusion and thus equilibration. To address this issue, it is helpful to first develop a solid understanding of how a synthesis of intermetallic nanocrystals is governed by the thermodynamic and kinetic parameters. In the following sections, we present a simplified nanoscopic model to describe the fundamental thermodynamic and kinetic features involved in the transition from disordered alloys to intermetallic compounds. This model mainly concentrates on the energetic aspects in the disorder-to-order transition (or the so-called ordering process), where a randomly mixed alloy (in a metastable state) equilibrates into an atomically ordered intermetallic phase (in a thermodynamically stable state) through enhanced diffusion. Such a transition is key to the success of any wet-chemical synthesis of intermetallic nanocrystals.

### 2.1. Thermodynamic analysis

The stability of a nanocrystal is determined by its thermodynamic state, which is defined by its size, shape, composition, and crystal structure. In the discussions to come, we focus on the thermodynamic aspects of intermetallic nanocrystals, and in particular, the characteristic temperature and composition at which a system of atoms favors a well-defined atomic ordering over another that is disordered. To begin, we first consider the equilibrium conditions of disordered and ordered phases in a bulk binary system (denoted as A-B). The thermodynamic state of a given atomic configuration (*i.e.*, disordered *vs*. ordered) can be defined by the Gibbs free energy (*G*), which depends on the enthalpy (*H*), entropy (*S*), and temperature (*T*), as given by:

$$G = H - TS \tag{1}$$

The equilibrium conditions for a bimetallic system are best summarized in a phase diagram, as shown in Figure 1a. The colored regions in the diagram represent the equilibrium phases as a function of composition (*x*) and temperature (*T*), with the ordered phase colored in orange and the disordered in blue. While these regions represent the most favorable phases in terms of thermodynamics, they do not provide any information with regard to the Gibbs free energies of the different phases. Such information can be found in Figure 1b, where the Gibbs free energy is plotted against composition along two isotherms, $T_1$ (dashed) and $T_2$ (solid), respectively. Along the high temperature isotherm ($T_1$ is supposed to be higher than the order-to-disorder transition temperature), the disordered solid solution is more stable than the ordered phase over the entire composition range. This is due to the higher entropy in the disordered atomic arrangement, which dominates the total free energy at high temperatures. When the temperature is sufficiently low, as along the low temperature isotherm ($T_2$), it is the enthalpy term that dominates the free energy contribution and the stronger bonding between A and B gives rise to a thermodynamically favored, ordered phase. The composition ranges over which the ordered phase is energetically favored at $T_2$ is marked by a red bar in Figure 1a.

It should be pointed out that the phase boundaries for solids and liquids (melts) are not given in this diagram for simplicity, and it is possible that a disordered phase region may not exist with a specific combination of elements and stoichiometry. In this case, an ordered intermetallic phase equilibrates with molten liquid, which not only is disordered in terms of elemental arrangement also lacks a lattice structure. In real material systems, one ordered intermetallic phase can equilibrate with another intermetallic phase when there are multiple present, and three-phase equilibrium can also occur (*e.g.*, the Sn-Pt and Zn-Pt systems). The basic A-$A_mB_n$-B relationship described in this chapter serves as the fundamental for other intermetallic equilibria that can be derived. However, caution should be taken when analyzing other material systems with different phase diagrams.

The thermodynamic driving force toward the ordered state from the disorder state in bulk is determined by the change in Gibbs free energy ($\Delta G_{d \to o}^{\infty}$), where $\Delta H_{d \to o}$ and $\Delta S_{d \to o}$ are the enthalpy and entropy differences between the disordered and ordered phases, respectively, as given by:

$$\Delta G_{d \to o}^{\infty} = \Delta H_{d \to o} - T \Delta S_{d \to o} \qquad (2)$$

The higher bonding energy in intermetallics (relative to the randomly mixed alloy) guarantees that $\Delta H_{d \to o}$ is negative for stable intermetallic configurations. However, since the system becomes more ordered, the change in entropy, $\Delta S_{d \to o}$, is also negative. As such, $\Delta G_{d \to o}^{\infty}$ is dominated by $\Delta H_{d \to o}$ at low temperatures, leading to a favorable disorder-to-order transition. At higher temperatures, in contrast, $\Delta G_{d \to o}^{\infty}$ is determined by the term $-T\Delta S_{d \to o}$, resulting in a favorable transition to the disordered phase. Conventionally, the temperature at which $\Delta G_{d \to o}^{\infty}$ = 0 is called the *critical phase transition temperature* ($T_{d \to o}^{\infty}$). In general, the magnitude of $\Delta G_{d \to o}$ can also have a major impact on the transition kinetics, which is discussed in detail in Section 2.2.

As for nanocrystals, the surface free energy plays a considerable role in the total free energy of the system. Therefore, to represent the change in Gibbs free energy for a given nanocrystal, a surface term must be added, as given by:

$$\Delta G_{d \to o}^{nano} = \Delta H_{d \to o} - T\Delta S_{d \to o} + \Delta \gamma_{d \to o} A \qquad (3)$$

where $\Delta \gamma_{d \to o}$ is the change in specific surface free energy between the disordered and ordered phases and $A$ is the surface area of the nanocrystal. Assuming that the shape does not change and thus $A$ remains constant during the phase transition, we can establish Equation (4) to describe how the equilibrium transition temperature in nanocrystals shifts relative to that of the bulk, where $T_{d \to o}^{nano}$ and $T_{d \to o}^{\infty}$ denote the critical temperatures for the disorder-to-order transition in nanocrystals and bulk materials, respectively.

$$\frac{T_{d \to o}^{nano}}{T_{d \to o}^{\infty}} = 1 + \left(\frac{A}{V}\right) \Delta \gamma_{d \to o} / \Delta H_{d \to o, V} \qquad (4)$$

The change in volume-specific bulk enthalpy ($\Delta H_{d \to o, V}$) is negative, with the same sign of $\Delta H_{d \to o}$. Accordingly, the change in specific surface free energy $\Delta \gamma_{d \to o}$ is positive because creating an ordered surface requires more energy than a disordered one because of the higher average bond energy. Since the surface-to-volume ratio increases with decreasing size, smaller nanocrystals should have lower disorder-to-order phase transition temperatures according to Equation (4). Furthermore, if we look at Equations (3) and (4), we can begin to infer the roles played by the composition, size, and shape in determining the favorability of intermetallic formation in nanocrystals.

*2.1.1. The effect of composition*

Different from alloys, intermetallic phases are only favored in a limited number of bimetallic systems that satisfy the thermodynamic demand for a sufficiently negative $\Delta H_{d \to o}$. As such, it is worthwhile to screen bulk phase diagrams to identify candidate metal pairs for intermetallics. Apart from simply analyzing bulk phase diagrams, suitable differences in electronegativity ($\chi$) and atomic radius can also be used to predict the possibility of forming intermetallics. This rule is based on the fact that these two parameters ultimately define the energies of A-A, B-B, and A-B bonds, which in turn defines $\Delta H_{d \to o}$. In general, large differences in electronegativity tend to contribute to covalent bonding. This reduces the miscibility of the two components, while it can also greatly lower the enthalpy of formation for the intermetallic phase (*i.e.*, leading to high structural stability). For example, Zn has the lowest electronegativity among its neighbors (*e.g.*, Fe, Co, Ni, and Cu). Accordingly, the Pt-Zn system exhibits stable and narrow intermetallic compositions rather than relatively broad regions of ordered phases, as summarized by the Hume-Rothery Rules —a set of useful guidelines for predicting the formation of alloys and intermetallic phases.[101]

When the stoichiometric proportion of a given bimetallic system deviates slightly from the favored atomic ratio, local inhomogeneity and some degree of atom disorder is expected. This alteration in chemical configuration leads to a further drop in $T_{d \to o}^{nano}$. In addition, phase segregation can occur at the surface of nanocrystals when the difference in surface free energy between the two metals is high enough.[102] As a system with a confined volume and a defined number of atoms, a nanocrystal that undergoes phase separation may suffer from a partial deviation from the preferred stoichiometry. For instance, Monte Carlo simulations suggest that the formation of a Pt-rich (001) surface and a smaller intermetallic core could lower the total free energy.[103] Moreover, surface segregation would be more prevalent when the surface-to-volume ratio is increased, leading to a completely disordered atomic arrangement. Surface segregation of Pt is often observed during the synthesis of Pt-M alloys and intermetallic nanocrystals, where M is usually a 3*d* transition metal.[104] In order to counteract this adverse deviation in composition, it may be necessary to adjust the feeding ratio of the different metal precursors during a synthesis.[105]

*2.1.2. The effects of size and shape*

The total surface free energy increases dramatically as the size of a nanocrystal approaches down to a few nanometers, resulting in the rapid drop in the disorder-to-order transition temperature.[106] This effect has been repeatedly confirmed in various bimetallic systems such as Pt-Co, Pt-Fe, and Au-Cu with the assistance of *in situ* heating inside a transmission electron microscope (TEM).[107-110] For example, Alloyeau and coworkers found that the phase transition temperature of PtCo nanocrystals with sizes of 2.4–3 nm was between 325–175 ºC lower than that of bulk.[111] In wet chemical syntheses, the nuclei of bimetallic nanocrystals are extremely small in size (< 1 nm), so the transition temperature can be as low as the reaction temperature. As such, an additional driving force is needed to promote atomic ordering during growth, which is often achieved through the restriction of kinetic barriers (see Section 2.2 for a discussion). For this reason, appropriate experimental conditions should be carefully selected to facilitate the formation of intermetallic nanocrystals with the desired size, shape, and stoichiometry.

In addition to size, the shape of a nanocrystal also plays an important role in determining the total change in Gibbs free energy. In most experiments, as well as corresponding theoretical treatments,[112-114] the identity of all the exposed facets is generally negated. However, the specific surface free energy of the individual facets must also be taken into consideration since it has non-negligible influence on the disorder-to-order transition temperature. Recently, changes to the ratio of facet area, for example, between {100} and {111}, were observed during the *in situ* heating of an individual $Pt_3Co$ nanocrystal.[115] Upon heating, it was found that the {111} facets became dominant when the ordered intermetallic phase was formed, as shown in Figure 2a. From the viewpoint of thermodynamics, the change in surface energy must also be considered when discussing the energetics of the disorder-to-order transition. This knowledge can also be employed to promote the formation of intermetallic nanocrystals by using capping

WILEY-VCHagents in wet chemical syntheses.

## 2.2. Kinetic analysis

Even though the experimental conditions may favor the intermetallic structure, complete transformation is typically never guaranteed within reasonable time frames due to the relatively high diffusion barriers. The kinetics of the transition process depends primarily on the competition between the thermodynamic driving force and the height of kinetic energy barriers that a system must overcome to reach the thermodynamic minimum, as illustrated in Figure 3a. The aim of this section is to identify the energy barriers that are involved in the synthesis of intermetallic nanocrystals, and to discuss the key factors that affect the kinetics of the disorder-to-order transition.

A comprehensive understanding of the nucleation kinetics of an ordered phase has been established by several papers based on the Johnson-Mehl-Avrami (JMA) theory.[116-118] In our analysis, a simplified kinetic model is adopted, where the two basic processes involved in a disorder-to-order transition are discussed separately, namely, the formation of a new phase within the parent phase (nucleation) and material transport (diffusion). It is worth noting that the underlying physics of these two processes, associated with bond breaking and formation, as well as atom diffusion, is the same. Similar to crystallization from a glassy state (the most commonly studied disorder-to-order transition),[119] the temperature-dependent reaction rate $R(T)$ can be described as the product of two components in Equation (5):

$$R(T) = f(T)D(T) \tag{5}$$

where $f(T)$ is the component representing the formation of the new phase and $D(T)$ represents the transport component. In most cases herein discussed, $R(T)$ exhibits a volcano-like profile due to the opposite dependences of $f(T)$ and $D(T)$ on temperature as shown in Figure 3b.

A few major assumptions and simplifications are made in the following analysis of the two rate terms in order to identify and analyze different factors on kinetic transition. First, we only consider the formation of a single new phase in a nanoparticle, regardless of their coarsening. Second, vacancy diffusion is considered as the dominant mode by which atoms move through a volume. Simple lattice jump is used to describe the diffusion in metals, assuming a dilute defect concentration and no interaction between the defects. These assumptions do not vary the qualitative trends derived from the kinetic analyses, so the general conclusions remain valid.

The formation rate of a new phase can be described by the Arrhenius relation, as given by Equation (6). The nucleation work ($W^*$) for the formation of an ordered phase within an alloy nanocrystal can be represented by the sum of the change in free energy, $\Delta G_{d \to o}^{nano}$ and the change in free energy associated with the formation of a disorder/order interface, $\gamma_{int} A_{int}^*$. Here, $\gamma_{int}$ and $A_{int}^*$ denote the specific interfacial free energy and the area of the interface, respectively.

$$f(T) \sim e^{-W^*/kT}, \qquad W^* = \Delta G_{d \to o}^{nano} + \gamma_{int} A_{int}^* \qquad (6)$$

As in the aforementioned thermodynamic analysis, the change in free energy $\Delta G_{d \to o}^{nano}$ is usually negative at low temperatures, and thus serves as the main driving force in most cases. At the same time, the term describing the change of surface energy is typically positive, presenting a size effect that lowers the driving force for the conversion of small nanocrystals. In addition to the extra obstacle associated with size-dependent surface free energy, the newly generated interfacial energy generally serves as a major energy barrier against the formation of a new phase.

Local atom rearrangement is a kinetically controlled process that facilitates phase transitions. The rate of atom movement within a lattice, that is, the atom transport rate, $D(T)$, can be calculated by considering random atomic jumping in Equation (7).[119-122] The directionality of diffusion or difference in chemical potential, on the other hand, is taken into account by the driving force term, $f(T)$.

$$D(T) \sim nvze^{-\Delta g_m/kT} \quad (7)$$

where the difference in chemical potential is taken into account in the driving force term. In Equation (7), the overall atom transport rate is associated with a lattice vibration model where $n$ is the concentration of the diffusive species and $v$ is the thermal vibration frequency. Based on the Einstein model where atoms vibrate as harmonic oscillators at lattice sites, $v$ is positively related to the bonding strength and negatively related to the effective mass of the atom. The number of the nearest neighbors for a diffusing species $z$ is typically the same as the coordination number in a specific crystal structure. In the exponential component $e^{-\Delta g_m/kT}$, $\Delta g_m$ is the energy barrier for an atom to jump from one site to another. In general, vacancy diffusion is considered to dominate the diffusion process for many metals because it has a much lower $\Delta g_m$ relative to the atom-exchange mechanism. For a vacancy-mediated process, the concentration of this kind of point defect can be increased through thermal activation, also following an Arrhenius relation $n \sim e^{-\Delta g_d/kT}$, where $\Delta g_d$ is the activation energy. Generally speaking, weakly bonded atoms will have both a lower defect formation energy and jumping barrier, which greatly accelerates the atom transport process. In contrast to the influence on driving force, raising temperature will effectively speed up the atom transport process since the signs of both $\Delta g_m$ and $\Delta g_d$ are positive. In summary, the composition, size, and shape are some of the key factors that will affect the formation of intermetallic nanocrystals by contributing to either the driving force or energy barrier.

*2.2.1. The effect of composition*

In general, the difference in crystal structure as well as other fundamental properties (*e.g.*, bonding strength) between the disordered and ordered phases are crucial factors that determine the driving force for a given phase transition. As noted previously, the surface energy of the interface, $\gamma_{int}$, is often positively related to the degree of structural changes occurring between the two phases, which mainly depends on the specific binary system. Taking the A1 structure of noble metals as an example, the similar or even same crystal structure between the two phases corresponds to a lower interfacial energy and larger driving force, usually facilitating a disorder-to-order transition process, such as the transitions into a $L1_0$ or $L1_2$ structure (space group: $Pm\bar{3}m$, but can be regarded as *fcc* when the lattice points are supposed to be equal). In contrast, a disorder-to-order transition involving a large change in crystal structure, for example, into the B2 structure in a body-centered cubic lattice (*bcc* when the lattice points are supposed to be equal; space group: $Pm\bar{3}m$), might need to overcome a higher energy barrier. As such, there are a considerable number of reports on the successful synthesis of intermetallic nanocrystals with the $L1_0$ or $L1_2$ structure, even at small sizes for PtFe and $AuCu_3$, while there are very few reports regarding intermetallic nanocrystals with the B2 structure, even at large sizes (*e.g.*, PdCu).[123-125]

As mentioned previously, disorder-to-order is a kinetic process with transport component $D(T)$ dominated by $\Delta g_m$ associated with the average bond strength of the parent phase. In general, the bond strength is related to the melting point, which provides good information with regard to the transition kinetics.[126] For a binary system with a relatively weaker average bond strength, the lower $\Delta g_m$ can be expected to expedite the transport process. For example, the efficient phase transition in Au-Cu systems benefits from the fast diffusion, which is in contrast to Pt-based systems that have relatively higher melting points.[127,128] On the other hand, the weak bond strength can facilitate the generation of vacancies in the crystal lattice through thermal activation due to the decreased $\Delta g_d$, which also promotes the diffusion kinetics. In addition to thermal activation, vacancies can be artificially introduced into the parent phase to accelerate the atom transport process. To this end, phase segregation of Au out from a PtFeAu alloy was used to produce additional vacancies and thus promote the transition into $L1_0$ PtFe@Au core-shell nanocrystals.[129] Beyond vacancies, other defects such as twin boundaries may also facilitate the atom diffusion process.[130]

*2.2.2. The effects of size and shape*

Different from thermodynamics, the effect of size on the kinetics of the disorder-to-order transition is more complicated, generally including two opposite impacts. The increase in surface energy will reduce the driving force toward an ordered phase whereas surface diffusion can make a greater contribution to the atom transport in smaller nanocrystals, with a much lower diffusion barrier to overcome relative to the bulk. Taken together, smaller nanocrystals facilitate a faster atom transport but restrict the formation of new phases. As an empirical rule, the adverse effect on the driving force is frequently dominant in this competition for very small nanocrystals (usually on the order of 1 nm), while the benefit of fast diffusion takes the lead at relatively larger sizes. As a result, there probably exists an optimal size range that favors the formation of intermetallic nanocrystals, which has been demonstrated in semiconductor nanocrystals.[131,132] This demonstration was also supported by a recent study on the seed-mediated co-reduction growth of B2 PdCu intermetallic nanocrystals (see Section 3.3 for a detailed discussion).[133]

In addition to the effect of size, shape (or faceting) can also have a profound impact on the transition kinetics. First, since different facets have distinctive surface and/or interface free energies, each will have a different driving force for the disorder-to-order transition according to Equation (6). This effect was clearly demonstrated for $Pt_3Co$ nanocrystals upon heating *in situ* under TEM.[115] In this study, the driving force on {110} facets was much larger than on {100}, leading to the preferred nucleation on {110} surfaces (see Figure 2b). Second, surface diffusion rate across different facets may also play a role in the preference of surface nucleation on a specific facet. However, no reports have been found to explicitly address this factor

In a typical synthesis, both the size and composition of nanocrystals change throughout the various stages of formation. Such dynamics is not captured by the aforementioned simple model, and to make matters more complicated, the dynamics of formation is supposed to be unique to each synthetic protocol. In principle, the size and composition must be optimized in order to produce an intermetallics in the growing parent phase to satisfy the thermodynamic and kinetic criteria for the transition. For this purpose, some diffusion-related physiochemical processes such as inter-particle diffusion, mixing of two elements (alloying), and particle overgrowth have been intensively explored for the synthesis of intermetallic nanocrystals. Compared to the atom transport in the local rearrangement process, the diffusion modes in the aforementioned processes are ruled by a similar kinetic principle but presented in different forms and efficiencies, which significantly affect the synthetic conditions in different ways. For example, slow inter-particle diffusion is the main approach used to grow the particles in the thermal annealing conversion from a disorder phase, which requires high temperatures and/or long time scales. On the contrary, fast surface diffusion in combination with particle overgrowth is dominant in phase transitions by the wet-chemistry approach (*e.g.*, one-pot synthesis), where intermetallic nanocrystals can be achieved at a mild temperature. Moreover, the intermixing (alloying) process within two separated solid phases (*e.g.*, thermal conversion from a separated phase) or from soluble species (*e.g.*, seed-mediated diffusion growth) contain both inter-particle and intra-particle diffusion, which display great differences in terms of diffusion rate for various binary systems. Overall, to rationally design synthetic protocols for intermetallic nanocrystals, we must first identify these specific fundamental physiochemical processes together with their corresponding diffusion modes.

## 3. Synthetic approaches to intermetallic nanocrystals

Bulk intermetallic compounds are typically produced through techniques such as powder metallurgy, arc melting, or induction heating, which are strategies reliant on equilibration through high-temperatures.[134-136] However, intermetallic catalysts with nanometer-sizes can be hardly obtained by these means, even after post-treatments such as crushing or ball milling.[137-139] Fortunately, the rapid development of colloidal chemistry in recent years has laid a solid foundation for the exploration into the formation of intermetallic nanocrystals from the bottom-up routes. The two primary steps include: (1) forming nanocrystals containing metallic elements as the nanometer-sized solid intermediates, and (2) converting these as-formed intermediates into the intermetallic phase without remarkably increasing the size. In this chapter, we will focus on the second step, during which the intermetallic nanocrystals are actually formed, and categorize the reported synthetic protocols based on the respective experimental conditions. In general, there are two major categories that include a thermal annealing approach performed in an atmosphere (or vacuum) and a wet-chemistry approach carried out in the solution phase. As summarized in Figure 4, these two major approaches can be further broken down into smaller subcategories based on the nature of the solid intermediates (*e.g.*, alloy or heterogeneous nanocrystals) formed in the first steps.

Generally speaking, the conversion from solid intermediates into intermetallic nanocrystals by thermal annealing in an atmosphere (or vacuum) is a direct and effective method that can be achieved at lower temperatures relative to those used in metallurgic approaches. However, elevated annealing temperatures (usually over 500 ºC) and/or long annealing times are sometimes required with this approach due to the relatively high kinetic barriers and corresponding low atomic mobility in the solid state. On the other hand, the height of the kinetic barriers involved in the atomic ordering (*i.e.*, alloy → intermetallic) are typically lower than those involved in the inter-diffusion process (*i.e.*, core-shell → alloy). This high-temperature annealing process usually causes the nanocrystals to aggregate, undergo Oswald ripening, and/or sinter as a result of their high specific surface free energy.[140-142] For nanocrystals initially in a far-from-equilibrium shape, the rapid surface diffusion enabled at elevated temperatures will cause the shape to quickly change in an effort to minimize the total surface free energy.[143] Such a transformation will also have an impact on the critical disorder-to-order transition temperature.

Wet chemical syntheses currently serve as the most versatile and powerful approach to the generation of bimetallic nanocrystals with well-defined compositions, sizes, shapes, and structures. In contrast to high-temperature annealing, heating in the solution phase inevitably involves various chemical species such as solvent molecules, capping agents, and reducing agents left over from the synthesis. These chemical species can potentially lower the surface free energy through surface chemisorption. According to the aforementioned kinetic analysis, lowering the surface free energy might result in a larger driving force for the disorder-to-order transition. By leveraging this effect, in conjunction with the enhanced surface diffusion, intermetallic nanocrystals can be produced using wet-chemistry approaches at relatively low temperatures. In this way, the reaction temperatures necessary for solid intermediate formation and their conversion into intermetallic phases can be tuned at very similar values, in which these two processes take place successively or even simultaneously. In general, since there is often a large difference in reduction potential between the two metal precursors involved, a strong reducing agent is necessary. This requirement can hinder the shape control of intermetallic nanocrystals owing to the rapid nucleation and growth. Meanwhile, the boiling point of a solvent severely limits the range of reaction temperatures, often resulting in a partial conversion for the disordered-to-ordered phase transition.

In spite of the aforementioned difficulties, a large number of synthetic protocols have been developed for generating various types of intermetallic nanocrystals. In the upcoming sections, we will discuss the recent progress and nuance features of these strategies.

### 3.1. Thermal annealing approach

According to the thermodynamic and kinetics analyses, the atomically disordered alloy exists in a metastable state relative to the intermetallic phase when maintained below the disorder-to-order transition temperature. If the disordered alloy nanocrystal is annealed at an elevated temperature for a sufficient period of time, it will be converted to an intermetallic nanocrystal with an ordered atomic arrangement.[144,145] Using this protocol, Sun and coworkers reported size-controlled synthesis of chemically ordered $L1_0$ PtFe intermetallic nanocrystals with high magnetocrystalline anisotropy by heating the disordered A1 alloy phase to 560–600 ºC under $N_2$ atmosphere (Figure 5).[146] The uniform PtFe alloy nanocrystals with sizes between 3–10 nm were prepared by simultaneously reducing $Pt(acac)_2$ and decomposing $Fe(CO)_5$ in the presence of oleylamine (OAm) and oleic acid (OA), with 1,2-hexadecanediol serving as the reducing agent. The homogeneous nucleation process was followed by seed-mediated overgrowth for the formation of PtFe alloy nanocrystals with larger sizes. This strategy has also been applied to the synthesis of intermetallic nanocrystals with different components and structures, such as PtZn, PtMn, and PtCo.[147-149] In these syntheses, size control was achieved by varying the size of the alloy nanocrystals and by optimizing the annealing temperature and time. The crystal structure of the intermetallic nanocrystals was determined by the composition of the alloy nanocrystals.

A large number of studies have demonstrated that conversion from alloy nanoparticles by annealing is the most effective method for synthesizing intermetallic nanocrystals, regardless of how the alloy nanocrystals were produced.[150,151] However, during the annealing process, the nanocrystals were prone to aggregation, Oswald ripening, and/or sintering, which inevitably leads to larger particles and broader size distributions.[152] This issue can be addressed by first dispersing the nanocrystals on an inert substrate such as carbon or silica to help immobilize them throughout the heating process and thus inhibit sintering.[153] However, confining structures to a two-dimensional surface ultimately reduces the overall throughput. As an alternative, alloy nanocrystals can be encapsulated in a protective shell prior to annealing. In one example, Korgel and coworkers demonstrated this method by first coating PtFe and $Pt_3Fe$ alloy nanocrystals with silica while suspended in water-in-oil microemulsions.[154] It was found that the silica shell could prevent the nanocrystals from sintering at annealing temperatures up to 850 °C due to the elimination of inter-particle diffusion. Free-standing $L1_0$ PtFe and $L1_2$ $Pt_3Fe$ intermetallic nanocrystals with fully ordered atomic arrangement were obtained after annealing in a reducing $H_2$/Ar gas, following by removal of the silica shell by etching in an alkaline solution. In another study, Sun and coworkers demonstrated the synthesis of dispersible $L1_0$ PtFe nanocrystals with an intermetallic phase by thermal annealing above 700 °C using MgO as a protective shell.[155] However, the intermetallic nanocrystals were partially ordered owing to the confined mobility of the two components in the MgO shell.[156] Recently, the same group solved this problem by annealing dumbbell-like $PtFe-Fe_3O_4$ nanocrystals in $H_2$/Ar gas, where ordering of the Fe and Pt was promoted by the defects generated during the reduction of $Fe_3O_4$ to Fe.[157] Interestingly, DiSalvo and coworkers employed a versatile method, pioneered by Armbrüster's group for the synthesis of PdGa intermetallics,[84] to avoid particle aggregation in the reaction and were able to control the coalescence during thermal annealing for the synthesis of intermetallic nanocrystals with various components.[158-161] The protocol involved the generation of an insoluble KCl byproduct upon rapid reduction of chloride-based precursors by a strong reducing agent such as $KEt_3BH$ in tetrahydrofuran (THF), which served as a protecting matrix for the formation of intermetallic nanocrystals in a subsequent thermal annealing treatment.

In addition to size and uniformity, there are strong efforts to develop shape-controlled syntheses. To this end, a simple strategy that one may propose would involve thermal annealing of alloy nanocrystals with well-defined shapes. However, the annealing temperatures required for the formation of most intermetallic phases in gas are typically higher than 500 °C, especially for attaining a fully ordered atomic arrangement, which would destruct the shape owing to the acceleration of surface diffusion. For example, Sun and coworkers evaluated the possibility to generate $L1_0$ PtFe intermetallic nanocrystals from alloy nanocrystals with cubic and wire-like shapes through thermal annealing.[162,163] Due to the rapid surface diffusion of atoms at numerous low-coordination sites (*e.g.*, corners and edges), the original shapes were quickly lost. In general, it is necessary to use relatively low annealing temperatures in order to preserve the shape during such a conversion process, making this strategy effective for certain systems. For example, it has been reported that bimetallic nanocrystals made of Au and Cu could be driven to form an ordered structure from a random, alloy phase at a relatively low annealing temperature below 400 °C.

Overall, the application of thermal conversion from shape-controlled alloy nanocrystals to bimetallic nanocrystals with an intermetallic structure has received limited success until recently. It is critical to use relatively low annealing temperature in order to preserve the shape of the alloy template, while at the same time allowing for the disorder-to-order transition. Unfortunately, this critical parameter is extremely restrictive in terms of the choice of components that are capable of producing intermetallic nanocrystals.

Bimetallic nanocrystals with an intermetallic structure can also be obtained by annealing heterogeneous nanostructures that are comprised of pure distinct metal phases that are connected at a common interface (*e.g.*, core-shell, dimer structure). In this strategy, heterogeneous structures are fabricated using wet chemical methods and then annealed at an elevated temperature under vacuum or in a reductive atmosphere to convert the biphasic structures into a homogenous intermetallic phase. In one example, Cheon and coworkers reported the synthesis of Co@Pt core-shell hollow nanocrystals of ~6 nm in size through galvanic replacement between Co nanocrystals and a salt precursor to elemental Pt.[164] This core-shell structure was subsequently converted to $L1_0$ PtCo intermetallic solid nanocrystals *via* thermal annealing under vacuum at 600–700 °C for 12 h. In another study, Adzic and coworkers found that incorporating Au into the shell of Co@Pd core-shell nanocrystals (*e.g.*, Co@PdAu) promoted their structural ordering at elevated temperatures.[165] When the Co@Pd and Co@AuPd nanocrystals were annealed at 800 °C under $H_2$, only the latter were converted to PdCo nanocrystals with intermetallic structures. The success of this synthesis could be attributed to the enhanced diffusion that occurred due to vacancy generation occurring after Au atoms segregated out from the AuPd alloy and to the surface of the nanostructures.

Apart from the use of systems comprised of two metals, this method can also be extended to metal oxides or hydroxides as precursors to the metals. This extension allows one to simplify the synthetic protocol by working with more diversified templates, and it is particularly useful for the synthesis of the intermetallics involving the active metals. For instance, Yang and coworkers successfully fabricated monodispersed $L1_0$ PtFe intermetallic nanocrystals of ~17 nm in size through thermal conversion from Pt@$Fe_2O_3$ core-shell nanostructures under a flow of $H_2$/Ar at 550 ºC.[166] In this synthesis, the $Fe_2O_3$ shell not only provided the Fe source but also acted as a protecting layer to prevent the particles from sintering. In another example, layered double hydroxides (LDHs) such as $Ni_2AlIn$-LDHs and $Ni_2Mg_7In_3$-LDHs were found to undergo topotactic transformation when calcined at 900 ºC in a reductive $H_2$/$N_2$ stream, resulting in metal oxides decorated with intermetallic nanocrystals made of Ni and In.[167] In this case, the LDHs served two functions, by acting as the source to both metal components and also as the precursor to the metal oxide support in the intermetallic nanocrystals. This system offers a simple and effective approach to the synthesis of intermetallic nanocrystals uniformly dispersed on a metal oxide support.

More generally, conventional oxide-supported monometallic catalysts, especially noble-metal nanocrystals, can be converted into intermetallic compounds at elevated temperatures in a reductive atmosphere, which is able to upgrade the catalytic performance. Armbrüster and co-workers generalized this phenomenon as "reactive metal-supporting interaction (RMSI)", which advanced the understanding of the physiochemical processes occurring during catalytic reactions.[168] As a typical example, the structural conversion of Pt/$CeO_2$ catalysts reduced at temperatures ranging from 200 to 950 ºC in $H_2$ was demonstrated using high resolution transmission electron microscopic (HRTEM) technique.[169] Figure 6a schematically exhibits the structural evolution processes undergone by the Pt/$CeO_2$ catalysts and the corresponding states of the catalyst are demonstrated by the HRTEM images (Figure 6b–d).

Taken together, a broad range of preformed heterostructures rather than alloy nanocrystals can be converted to intermetallic nanocrystals through thermal annealing in a reducing atmosphere or in vacuum. As a major advantage, this strategy can potentially improve the dispersion of intermetallic nanocrystals as a result of the protective oxide shell. However, due to the substantial changes in volume, shape, structure, and composition during the conversion process, it is still challenging to apply this method to the synthesis of intermetallic nanocrystals with controlled shapes. In addition, the ordering process often requires thermal annealing for a relatively long period of time because of the slow inter-diffusion rates associated with such heterostructured systems.

### 3.2. Wet-chemistry approach

In order to avoid severe aggregation at elevated temperatures during conversion processes, wet chemical methods have received increasing interest for the production of well-dispersed intermetallic nanocrystals in the presence of stabilizers at relatively low temperatures.

Similar to the aforementioned cases in the gas phase where the conversion was reliant on atomic inter-diffusion,[170,171] seed-mediated diffusion growth has also been established as a viable method for generating intermetallic nanocrystals, but in the solution phase. This protocol generally involves two sub-steps: *i*) synthesis of seeds in the form of metal nanocrystals, and *ii*) diffusion of atoms from the newly generated species including atoms, clusters, and small particles into the as-formed seeds.

This synthetic strategy was pioneered by Schaak and coworkers, who were among the first to report a general strategy for generating intermetallic M-Zn nanocrystals (M = Au, Cu, and Pd) through seed-mediated diffusion growth using zero-valent organometallic Zn precursors in hot organoamine solvents.[172] In such a synthesis, the size, shape, and internal structure of the final products can be tuned to a certain extent by employing seeds with a specific twin structure. To this end, Li and coworkers demonstrated the synthesis of uniform 5-fold twinned AuCu intermetallic nanocrystals of ~10 nm in size by employing pseudo-icosahedral or -decahedral Au nanocrystals as the seeds in a mixture of OA and tri-n-octylamine at 280 °C (Figure 7a).[173] In this synthesis, the newly reduced Cu atoms or clusters with high reactivity rapidly diffused into the 5-fold twinned Au seeds as driven by the thermal energy, leading to the formation of $L1_0$ AuCu intermetallic nanocrystals. By increasing the amount of the Cu precursor, the $L1_2$ $AuCu_3$ intermetallic nanocrystals were generated at a higher reaction temperature of 300 °C, as shown in Figure 7b. Due to the strong capping of OA, the intermetallic nanocrystals were inhibited from anisotropic growth in this system. In another study, Ying and coworkers showed the anisotropic growth of the intermetallic AuCu pentagonal nanorods by replacing a mixture of OA and tri-n-octylamine with OAm.[174] The intermetallic AuCu pentagonal nanorods were preferentially grown from the well-defined, multiply-twinned decahedral Au seeds along their <110> directions due to the selective adsorption of OAm on the {100} facets. Using this strategy, the length, structure, and composition of the intermetallic AuCu pentagonal nanorods could be easily maneuvered by varying the number of Au seeds and their molar ratio relative to the Cu precursor, respectively (Figure 7, c–e). In another study, Jin and coworkers reported the synthesis of $Au_3Cu$ intermetallic truncated nanocubes with sizes of 15–30 nm through the chemical conversion of Cu microparticles in a mixture containing OAm, trioctylphosphine (TOP), and $AuPPh_3Cl$ at 200 °C (Figure 7f).[175] The synthesis was initiated by a galvanic replacement reaction occurring between the Cu microparticles and the Au precursor, leading to the formation of single-crystal Au nanocrystals and $Cu^{2+}$ ions. Afterwards, the newly formed $Cu^{2+}$ ions were reduced to Cu atoms by OAm, followed by diffusion into the single-crystal Au seeds to produce intermetallic truncated nanocubes. Interestingly, only $Au_3Cu$ intermetallics could be generated even when the Cu precursor was in excess because this phase is the most thermodynamically stable among three types of Au-Cu intermetallic compounds (*i.e.*, $Au_3Cu$, AuCu, and $AuCu_3$).[176]

The difference in diffusion rate between two components plays a critical role in determining the inner structure (*i.e.*, solid *vs.* hollow) of the final products, a physical process commonly referred to as the nanoscale Kirkendall effect.[177] This concept has been widely exploited for the formation of hollow intermetallics from core-shell nanocrystals. In essence, when the outward diffusion of the core metal is significantly faster than the inward diffusion of the shell metal, the vacancies that are left behind at the core will eventually coalesce to form a void, resulting in the formation of a hollow structure.[177,178] In the synthesis of intermetallic nanocrystals *via* seed-mediated diffusion growth, the Kirkendall effect can also be prevalent. For instance, Schaak and coworkers utilized tetragonal Sn nanocrystals as reactive templates to produce binary M-Sn (M = Fe, Co, Ni, and Pd) intermetallic compounds *via* reaction with the corresponding precursors in tetraethylene glycol (TEG), with $NaBH_4$ acting as a strong reducing agent.[179] For $FeSn_2$ (C16; space group: *I4/mcm*) intermetallic nanocrystals, Sn diffused much faster than Fe, leading to void generation through the Kirkendall effect. As such, the cubic Sn nanocrystals with sizes larger than 15 nm were converted to $FeSn_2$ intermetallic nanocrystals with various shapes including hollow squares, U-shaped structures, and nanorod dimers. The variation in shape arose from the anisotropic structure of the tetragonal Sn cubes, since the top and bottom faces had a lower density of surface atoms relative to the four side faces. This unique structural feature was responsible for the anisotropic diffusion of Fe into the cubes from the top and bottom faces and outward diffusion of Sn along different axes, leading to an anisotropic Kirkendall effect. The synthesis of $FeSn_2$ intermetallic nanocrystals *via* chemical conversion was also sensitive to the morphological features of the tetragonal Sn seeds due to the size and shape-dependent reactivity. In another example, Rioux and coworkers carried out a systematic study on the Kirkendall process for the formation of hollow NiZn (B2) intermetallic nanocrystals with sizes of ~12 nm through the chemical conversion of Ni nanocrystals using a Zn precursor in hot organoamine (Figure 8).[180] The authors found that unequal diffusion rates between the Ni and Zn atoms led to the generation of vacancies in the interior of the nanocrystals, followed by vacancy condensation and void formation at 10 min after the injection of Zn precursor. Subsequently, the void expanded rapidly and reached a maximum size of ~4 nm at 30 min. This work provides another example for the formation of hollow intermetallic nanocrystals induced by the nanoscale Kirkendall effect.

For syntheses involving single component metal nanocrystals as the seeds, the chemical conversion typically requires high temperatures to overcome the high diffusion barriers and relatively long diffusion distances. In some cases, the temperatures necessary for complete inter-diffusion cannot be reached. For example, Li and coworkers investigated the diffusion of Ni into highly branched Pt nanobundles in octadecylamine (ODA) at 250 °C, showing that the atomic ratio of Ni/Pt only reached 1:9 and the random alloys were eventually generated even when the Ni precursor was excessive.[181]

Compared to the transformation from seeds made of a single metal, the chemical conversion between intermetallics with different structures and compositions can be carried out more easily in solution owing to the shorter average diffusion length and also the lowlihood that the diffusion barrier is relatively lower. As a typical example, Schaak and coworkers demonstrated the reversible conversion among C1 (space group: $Fm\bar{3}m$) of $fcc$ PtSn$_2$, B8$_1$ (space group: P6$_3$/mmc) of hexagonal close packed ($hcp$) PtSn, and L1$_2$ Pt$_3$Sn by aging one type of intermetallic nanocrystals with the corresponding precursors in TEG at 255–280 °C.[182]

In addition to the use of intermetallic nanocrystals as seeds, there are some reports on the chemical conversion of bimetallic nanocrystals with an alloy structure into an intermetallic structure.[183] In a recent study, Skrabalak and coworkers demonstrated the synthesis of uniform PdCu intermetallic nanocrystals with a B2 phase by a seed-mediated co-reduction process using PdCu random alloy nanocrystals as the seeds in a mixture containing OAm and TOP at 270 °C for 30 min (Figure 9).[133] In this synthesis, the ordering process proceeded gradually with the overgrowth of the seeds, showing a dependence on the particle size. This observation can be attributed to the activation barrier arising from the size-dependent surface energies, which has a direct impact on the driving force to undergo the disorder-to-order transition. In addition, the authors found that the phase transformation in the seed-mediated co-reduction synthesis to be intra-particle diffusion dependent, leading to a shorter growth time as compared with thermal conversion approaches dependent on inter-particle diffusion.

As illustrated by the aforementioned examples, seed-mediated diffusion growth in solution is a powerful route to the synthesis of intermetallic nanocrystals with well-controlled dispersity, size, shape, composition, and structure at mild temperatures, which is facilitated by the choice of the seeds. The formation of intermetallic nanocrystals is size dependent since the surface energy and activation barrier for ordering trends with size, as discussed previously in Section 2. Manipulating the composition of the seed (*e.g.*, mono- *vs.* bimetallic) or atomic ordering (*e.g.*, random alloy *vs.* intermetallic structure) provides additional handles for tuning the structure over a broad range. In addition, the use of capping agents might activate the ordering process and thus facilitate the formation of intermetallic structures at mild temperatures by decreasing the surface energy of nanocrystals *via* strong adsorption. Moreover, a capping agent can selectively chemisorb onto specific types of facets to enable shape control.[184] In seed-mediated growth, kinetic control is also an effective strategy for maneuvering the shape of nanocrystals.[185] Combining the use of capping agents and kinetic control is expected to emerge as a highly versatile strategy for controlling the structure and shape of intermetallic nanocrystals.

Thanks to the relatively mild conditions required for the conversion in solution, both of the steps involved in the production of intermetallic nanocrystals can be achieved in some cases simply through a one-pot synthesis, which is arguably the simplest wet chemical route. In this case, co-reduction with a strong reducing agent such as $NaBH_4$, *n*-butyllithium, or sodium naphthalide, is often required due to the large difference in reduction potential between the two metal precursors involved.[186-189] In one example, Schaak and coworkers established a general approach to the low-temperature synthesis of phase-pure binary intermetallic nanocrystals with a large range of compositions (*e.g.*, AuCu, Sn-based, and Pt-based) in polyol (*e.g.*, TEG) using $NaBH_4$ as a strong reducing agent.[190-193] As other typical examples, intermetallic nanocrystals with compositions of PtBi or PtPb were obtained at room temperature by DiSalvo and coworkers when co-reduction was carried out with NaBH$_4$ in a methanol solution.[194,195] The same group also demonstrated the synthesis of PtPb and PtBi intermetallic nanocrystals using sodium naphthalide as a reducing agent in the boiling diglyme or THF solution.[188, 196]

In contrast to systems involving Au and Cu with complete miscibility, Au is known to show very limited miscibility with the ferromagnetic metals (*e.g.*, M = Fe, Co, and Ni), making their formation difficult under equilibrium conditions. Schaak and coworkers successfully generated the Au$_3$M intermetallic nanocrystals with a L1$_2$ structure by manipulating the reduction kinetics under non-equilibrium conditions.[197,198] This synthesis involved the rapid co-reduction of two metal precursors in octyl ether or diphenyl ether containing a small amount of OAm with n-butyllithium as an extremely active reductant at ~250 °C. Recently, Guo and coworkers developed a one-step co-reduction method for the synthesis of B8$_1$ PtBi and C2 (space group: *Pa3*) PtBi$_2$ intermetallic nanocrystals in a continuous-flow microfluidic reactor within seconds at 260 and 350 °C, respectively.[199] This method could be extended to produce other intermetallic nanocrystals made of Pt$_3$Fe or PtSn. For a system involving two components with relatively low electronegativities, the use of a strong reducing agent is essential. Most recently, Wang and coworkers demonstrated the synthesis of L1$_2$ Ni$_3$Sn, monoclinic Ni$_3$Sn$_4$, and hexagonal Ni$_3$Sn$_2$ intermetallic nanocrystals by co-reducing two metal precursors with different molar ratios in OAm using n-butyllithium as a reductant at 250 °C for 4 h.[200] The modified approach was also used to produce L1$_2$ Ni$_3$Ga and orthorhombic Ni$_5$Ga$_3$ intermetallic nanocrystals by replacing n-butyllithium with *tert*-butylamine-borane (TBAB), which served as the reducing agent. Despite the high efficiency in producing intermetallic nanocrystals *via* the co-reduction method using a strong reductant, shape-control of intermetallic nanocrystals in these syntheses is still challenging due to the rapid rates of nucleation and growth.

Maneuvering the shape of intermetallic nanocrystals conventionally requires the use of a mild reducing agent to decrease the reaction rate in a controllable manner, together with the addition of a specific stabilizer and/or capping agent to direct shape. Owing to the high oxophilicity associated with the low electronegativity of the elemental component in intermetallics, an oil phase and air-free synthetic environment is often required. To this end, Yang and coworkers demonstrated the synthesis of $B8_1$ PtPb intermetallic nanorods by simultaneously reducing Pt and Pb precursors with TBAB in diphenyl ether containing adamantanecarboxylic acid (ACA), hexadecanethiol (HDT), and hexadecylamine (HDA) at 180 °C under an inert atmosphere (Figure 10a).[201] The PtPb intermetallic nanorods were formed by the preferential overgrowth along the <001> direction due to the stronger capping of the top and bottom faces instead of the side faces with a hexagonal structure (Figure 10b). The combination of ACA, HDT, and HDA in a mixture was also responsible for the formation of PtPb intermetallic nanorods *via* selective adsorption and modification of the reaction kinetics. Yu and coworkers reported the production of PdCu intermetallic nanocubes with a B2 structure using OAm as both a solvent and a reducing agent in the presence of TOP as a stabilizer and capping agent at 250 °C (Figure 10c).[202] The use of TOP played an important role in facilitating the formation of the cubic shape by capping their {100} facets *via* selective adsorption of P-based groups derived from TOP (Figure 10d). Recently, Cabot and coworkers demonstrated the synthesis of orthogonal $Pd_2Sn$ nanorods (space group: Pnma) with different aspect ratios in OAm containing hexadecylammonium chloride (HDA·HCl) or methylamine hydrochloride (MAHC) with TOP as a stabilizer and capping agent at 300 °C (Figure 11).[203,204] The authors found that $Cl^−$ ions derived from HDA·HCl or MAHC could selectively desorb TOP from the {010} facets on the two ends of a nanorod, leading to anisotropic growth (Figure 11a). On the basis of this feature, the length and diameter of the $Pd_2Sn$ nanorods were easily tuned by varying the relative concentrations of TOP and $Cl^−$ ions (Figure 11, b–e). In another study, Hou and coworkers demonstrated the synthesis of PtBi intermetallic nanoplates with a $B8_1$ structure by simultaneously reducing Pt and Bi salt precursors in OAm at 200 °C with $NH_4Br$ as a capping agent (Figure 12a and b).[205] The nanoplates were formed due to the selective adsorption of bromide ions on the {101} facets of the anisotropic hexagonal structure. The authors found that the type of metal precursors and the use of other halide ions (*e.g.*, Cl⁻ and I⁻ ions) had a great impact on the shape of the nanoplates formed, by demonstrating the formation of square, disk, triangular, and hexagonal plates (Figure 12, c–f). Another example of intermetallic nanocrystals with a two-dimensional shape has been reported most recently by Huang and coworkers, who obtained PtPb@Pt nanoplates (Figure 13) with B8$_1$ PtPb core and A1 Pt shells of ~4 atomic layers in the OAm/octadecene mixed solution.[206] During the growth of intermetallic PtPb nanocrystals, Pb$_3$(CO$_3$)$_2$(OH)$_2$ was firstly formed followed by the transformation of this kind of precursor and the reduction as well as inter-diffusion of Pt. In addition, ascorbic acid (AA) was employed in the synthesis not only as a reducing agent but also as a weak acid to remove Pb atoms, which facilities the in-diffusion of Pt and rearrangement processes.

Recently, Li and coworkers developed a general method to the production of a large number of Au-, Pd-, Pt-, Ir-, Ru-, and Rh-based bimetallic nanocrystals with a homogeneous phase including those with intermetallic structures.[207] In this case, ODA simultaneously served as the solvent, surfactant, and reducing agent. An effective electronegativity rule was proposed as a guide to produce bi- and even tri-metallic nanocrystals with a given composition in the ODA system, where the components with an effective electronegativity larger than 1.93 could be co-reduced to form the homogenous phase. However, the formation of intermetallic or alloy nanocrystals was strongly dependent on the components or compositions. For example, only when the Cu and Pt precursors were fed into the synthesis with a molar ratio of 1 or 3, could the PtCu and PtCu$_3$ intermetallic nanocrystals be successfully produced, otherwise, alloy nanocrystals made of Pt and Cu were obtained. The variation in molar ratio of the two metal precursors also led to the shape evolution of the PtCu bimetallic nanocrystals due to the different reduction kinetics associated with their effective electronegativities. Wang and coworkers systematically investigated the shape and structure evolution of intermetallic nanocrystals consisting of Pt and Sn prepared in octadecene at 300 °C where dodecylamine (DDA) and 1,2-hexadecandiol (HDD) served as the capping agent and reducing agent, respectively.[208] In this study, the $Pt_3Sn$ intermetallic nanocrystals with a shape yield consisting of 85% cubes and 15% tetrahedra were fabricated at a nominal Pt/Sn molar ratio of 70:30 likely due to the selective capping of the amine group (Figure 14a). Decreasing the nominal Pt/Sn molar ratio to 40:60 resulted in the intermetallic structure evolution from $L1_2$ to $B8_1$, together with a shape change (Figure 14b). Different from the $Pt_3Sn$ sample, PtSn intermetallic polyhedra with a less defined shape were formed in the final product. At a nominal Pt/Sn molar ratio of 20:80, large faceted $PtSn_2$ intermetallic nanocrystals were obtained, which was attributed to the insufficient stabilizing ability of DDA when elevated amounts of Sn was used (Figure 14c). Interestingly, replacing DDA with a small amount of OAm and OA resulted in the formation of PtSn intermetallic nanowires *via* linear coalescence of small spherical nanocrystals driven by the reduction in overall surface energy (Figure 14d).

In one-pot syntheses involving the use of mild reducing agents, the large difference in redox potentials between the two metal precursors often results in the formation of monometallic (where the other precursor remains unreacted) or one-metal-rich nanocrystals preferentially. If one precursor is completely reduced first, the resulting seeds promote the formation of intermetallic nanocrystals by diffusion with the second species, which is similar to the seed-mediated diffusion growth (Section 3.3). This two-step process in a one-pot synthesis offers a great opportunity to tune the shape and surface structure of intermetallic nanocrystals using various strategies such as template directing, underpotential deposition (UPD), kinetic control, or galvanic replacement. To this end, Huang and coworkers reported the synthesis of crenel-like $Pt_3Co$ hierarchical nanowires with a $L1_2$ intermetallic structure, high-index facets, and Pt-rich surfaces in OAm at 160 °C for 8 h, where glucose and cetyltrimethylammonium chloride (CTAC) acted as the reducing and structure-directing agents, respectively (Figure 15).[209] In the initial stage, pure Pt nanowires of ~2 nm in diameter were produced *via* the reduction of the Pt precursors with glucose and then preferential Pt growth along the <110> direction. After that,

Pt₃Co intermetallic structures were formed through the inter-diffusion between Pt nanowires and newly formed Co species, which generated the hump-like sub-structures that grew perpendicular to the axial direction of the Pt nanowires. In this synthesis, successful production of the hierarchical Pt₃Co nanowires relied on the initial formation of thin Pt nanowires, which was controlled by the use of CTAC, Co(acac)$_3$, and glucose at the appropriate amounts.

It is a great challenge to generate intermetallic nanocrystals using a mild reducing agent in a system involving components with very large difference in standard redox potential, such as Pt and Zn (Pt$^{2+}$/Pt of 1.18 V vs. Zn$^{2+}$/Zn of -0.76 V versus RHE). To overcome this limitation, in a recent study, Xie and coworkers developed a UPD-based strategy to synthesize Pt₃Zn intermetallic concave cubes mainly enclosed by high-index {520} facets using N,N-dimethylformamide (DMF) as both the solvent and mild reducing agent in a sealed vessel at 180 °C for 9 h (Figure 16).[210] The UPD of Zn atoms on newly reduced Pt remarkably reduced the difference in standard reduction potential of the Zn and Pt precursors, facilitating the formation of a homogeneous Pt₃Zn phase under mild conditions. The moderate growth rate further ensured that the reduced Zn atoms had enough time to deposit and diffuse to positions in the Pt₃Zn intermetallic nanocrystals to achieve a thermodynamic minimum. As such, the type of Pt precursor used, together with the use of DMF, played an important role in the formation of the intermetallic structure since these experimental parameters greatly influence the growth rate. In addition, the UPD of Zn atoms on the specific facets of Pt is essential to the formation of concave cubes with high-index facets.

In addition to the UPD-based strategy, kinetic control represents a powerful and versatile approach to the synthesis of alloy nanocrystals or nanocrystals with a concave structure.[41, 211] As a model example, DiSalvo developed a kinetically controlled polyol process to obtain PtBi intermetallic nanocrystals in the absence of the additional reducing agent.[212] In this synthesis, the solution containing the Pt precursor was slowly added into a heated solution containing a Bi precursor that was partly pre-reduced to balance the notable difference in reduction potential between the precursors. Most recently, Li and coworkers demonstrated the solvothermal synthesis of cubic, concave cubic, and defect-rich cubic intermetallic $Pt_3Sn$ nanocrystals with a $L1_2$ structure by varying the concentrations of the Pt and Sn precursors under a kinetic control (Figure 17).[213] The synthesis involved the reduction of $Pt(acac)_2$ and $SnCl_2$ by DMF at 180 $^oC$ for 6 h. It was proposed that the oxidative etching of Pt caused by $Cl^-$ ions and $O_2$ assisted the diffusion of Sn atoms into Pt under the solvothermal conditions,[214] and thus facilitated the alloying and ordering processes in parallel with the overgrowth. The oxidative etching in combination with kinetic control were responsible for the formation of cubic $Pt_3Sn$ intermetallic nanocrystals with three types of surface structures encased by high-index facets, which could be tuned by varying the Pt and Sn precursor concentrations.[215] The kinetically-controlled and etching assisted method could be extended to synthesize concave and defect-rich nanocubes made of other intermetallics such as $Pt_3Mn$.

In the synthesis of $PtCu_3$ intermetallic nanocrystals, the use of chemical species containing amine group can be used to interact strongly with Cu precursors to promote the formation of decomposable Cu-based complexes. To this end, it was demonstrated that Cu-based complexes could be decomposed before the reduction of the Pt precursor, leading to the galvanic replacement between the newly formed pure Cu or Cu-rich nanocrystals and the Pt precursor.[217] By utilizing this feature, Lou and coworkers demonstrated a solvothermal approach to the synthesis of the $PtCu_3$ nanocages with an intermetallic structure using OAm as both the solvent and reducing agent in the presence of cetyltrimethylammonium bromide (CTAB).[218] In this synthesis, Cu nanocrystals were generated prior to the reduction of Pt regardless of its intrinsic lower standard reduction potential. The subsequent galvanic replacement between the preformed Cu nanocrystals and Pt species in combination with a co-reduction reaction eventually led to the formation of intermetallic $PtCu_3$ hollow nanocages.

In one-pot syntheses, the use of a relatively strong reducing agent shows great power in facilitating the formation of intermetallic nanocrystals, especially for systems containing components with low electronegativity, high bonding energy, or immiscibility in the bulk, probably due to the ease in initiating the alloying process. Unfortunately, shape control in these syntheses is yet to be achieved due to the uncontrolled nucleation and growth process associated with the rapid reaction rate. The use of mild reducing conditions and capping agents represent a promising strategy for the shape-controlled synthesis of intermetallic nanocrystals through preferential overgrowth, in particular for those with anisotropic structures. Significantly, various physicochemical processes such as UPD, kinetic control, and galvanic replacement often coexist in such a synthesis, which can be employed to further control the intermetallic nanocrystals with a concave or hollow structure.

In summary, the representative work on intermetallic nanocrystals with well-defined shapes is shown in Table 1, in which the crystal structures, shapes, exposed facets, and catalytic applications are included. Due to the limitation of thermal annealing approaches in shape-controlled synthesis of intermetallic nanocrystals, all of the representative intermetallic nanocrystals with well-defined shapes are produced by wet-chemistry strategies.

## 4. Catalytic applications of intermetallic nanocrystals

The design of advanced catalysts generally follows the Sabatier principle, in which the activity of a catalyst is determined by the interactions between the reaction species involved and the active sites on the catalyst. The binding strength between them should be optimized at an appropriate value in order to balance the adsorption and desorption strength and thus achieve the highest activity for a given reaction. According to Nørskov and coworkers, the binding strength is directly related to $d$-band center position (relative to the Fermi level) of a catalyst.[219] Density-functional theory (DFT) studies indicate that the position of the $d$-band center for a bimetallic catalyst has been found to depend on both electronic and geometric effects arising from the interaction between these two metals.[220-221] In intermetallic nanocrystals, relative to alloys with randomly ordered atoms, the well-defined composition and surface structure endow them with a unique and predictable electronic structure. Moreover, the electronic effect can be strengthened through the combination of localization of electron and directional covalent bonding. In addition, the change of bonding length and coordination number during the formation of intermetallic nanocrystals can also alter the position of the $d$-band center, and thus the catalytic activity (*i.e.*, geometry effect).[222]

In addition to catalytic activity, achieving high selectivity towards targeted products is a chief task in designing a catalyst. In general, high selectivity is when a reactant preferentially goes down a specific and desired reaction pathway that leads to the formation of the desired product by avoiding competing pathways that lead to unwanted products. According to site isolation concept, high selectivity can be realized by decreasing the homoatomic coordination number of active sites. This beneficial effect has been displayed in bimetallic catalysts, among which intermetallic nanocrystals are specifically desirable due to their unique surface structures. In addition, the more negative enthalpy of formation for intermetallics, relative to the random alloy phases, can be expected to enhance the chemical stability of intermetallic catalysts during catalytic processes.

Over the past decades, these aforementioned effects have been exploited to achieve catalysts with remarkably improved performance for a rich variety of catalytic reactions.[223] In this section, we present plenty of examples that highlight the excellent catalytic properties of intermetallic nanocrystals.

**4.1. Pt-based intermetallic nanocrystals for catalysis**

Nanocrystal catalysts comprised of Pt are widely used as electrocatalysts in industrial processes and commercial devices. Such catalysts are well-known for handling a wide variety of reactions, especially for ORR at the cathode and the oxidation of fuels (*e.g.*, methanol, ethanol, and formic acid) at the anode in proton-exchange membrane (PEM) fuel cells.[224-226] However, the sluggish kinetics of ORR arising from the strong bonding between Pt and oxygenated intermediates requires the loading of a large quantity of Pt per device, a feature that severely hampers the commercialization of this clean-energy technology since Pt is extremely scarce and costly.[227] To reduce the loading of Pt, researchers have investigated the effect of alloying Pt with transition metals such as Fe, Co, or Ni. Alloying has been shown to downshift the *d*-band center of Pt, a key feature that weakens the bonding between Pt surface atoms and oxygenated species, which leads to remarkably enhanced ORR activities.[228,229]

If the alloys are allowed to equilibrate into intermetallic nanocrystals, the electronic and geometry effects will be further strengthened and additional enhancements could be observed. To this end, Sun and coworkers compared the ORR properties of $L1_0$ PtFe intermetallics and A1 PtFe alloy nanocrystals to commercial Pt/C. They showed that the intermetallic nanocrystals were more active and durable relative to both the random alloys and commercial Pt/C.[156] They attributed the enhancement in durability to the superior chemical stability of the intermetallic Pt and Fe arrangement (*e.g.*, a small Fe loss of 3.3% for the intermetallics *vs.* a heavy Fe loss of 36.5% for the random alloys after immersion in 0.5 M $H_2SO_4$ for 1 h). In a later report, fully ordered PtFe intermetallic nanocrystals were produced by annealing MgO-coated dumbbell-like PtFe-$Fe_3O_4$ nanocrystals.[157] Compared to the partially ordered $L1_0$ PtFe, A1 PtFe, and commercial Pt/C, the fully ordered PtFe intermetallic nanocrystals exhibited the highest specific (3.16 mA/$cm^2$) and mass (0.69 A/$mg_{Pt}$) activities towards ORR at 0.9 V, which were 11.2 and 5.3 times higher than those measured from commercial Pt/C, respectively (Figure 18). Accelerated durability test (ADT) indicated that the fully ordered PtFe intermetallic nanocrystals displayed the best durability, with no obvious drop in activity even after 20000 cycles of ADT. The substantially enhanced ORR activity and durability could be attributed to the strong electronic and geometry effects between Pt and Fe and the enhanced resistance to acidic environments due to the fully ordered atomic arrangement, respectively. The mass activity could be further enhanced by 7.4 times (relative to the commercial Pt/C) by decreasing the particle size to 3.6 nm.[230] The enhancement in mass activity can be directly related to the increase in the electrochemical active surface area (ECSA).

Coating an intermetallic nanocrystal with a Pt skin also provides another promising strategy for enhancing the catalytic performance of nanocrystals towards ORR. There are two immediate advantages here, which include *i*) the reduction in Pt loading (the amount of Pt is drastically reduced since the skin is comprised of only a few Pt atomic layers) and *ii*) the favorable modulation of the *d*-band center (the electronic structure of the Pt skin can be tuned by the strong interaction with the underlying core).[231] Synthetically, the skin can be generated using a number of strategies including thermal segregation, dealloying, UPD followed by galvanic replacement, and seed-mediated growth.[232-235] To this end, Sun and coworkers utilized an electrochemical dealloying process to synthesize $L1_0$ PtFe and A1 PtFe nanocrystals encapsulated in a three-monolayer Pt skin having the same Pt/Fe atomic ratio of 75:25, where the corresponding intermetallic and random alloy nanocrystals were used as the templates, respectively.[236] The authors found that the $L1_0$ PtFe@Pt nanocrystals exhibited much higher specific activity (2.10 mA/cm$^2$) relative to the A1 PtFe@Pt catalysts (0.89 mA/cm$^2$) and commercial Pt/C (0.26 mA/cm$^2$). Theoretical calculations showed that the compressive strain existed in the Pt-skin layers of both structures, and the compressed strain can relieve the over-bonding with the oxygenated species by the downshifting the *d*-band center of the Pt atoms, thereby leading to an enhancement in the ORR activity. However, the Pt skin on the A1 PtFe@Pt nanocrystals was to be over-compressed, resulting in relatively lower ORR activity as compared to the $L1_0$ PtFe@Pt nanocrystals. Interestingly, if Fe atoms are partially substituted with Cu atoms, which have a larger atomic radius (~128 *vs*. 126 pm), to form $L1_0$ PtFeCu@Pt structures, the compressive strain can be partially relaxed and the highest specific activity (2.55 mA/cm$^2$) toward ORR can be achieved.

Recently, another great success in tuning the surface strain for boosting the ORR performance of Pt-based intermetallic catalysts was achieved in an unconventional manner. Huang and coworkers demonstrated a plate-like PtPb@Pt structure with {110} planes of Pt coherently grown on the {010} and {001} planes of $B8_1$ PtPb core, in which a very high tensile strain (~7.5%) and little compressive strain (~1%) were created along Pt <110>.[206] Through DFT calculation, the authors found that the tensile strain in the <001> direction could weaken the Pt-O binding on Pt{110} facets, thus, increase the overall ORR performance. This structure-performance relationship was further confirmed with the results of ORR tests. The PtPb@Pt nanoplates exhibited an ultra-high specific as well as mass activity at 0.9 V (vs. RHE) of 7.2 mA/cm$^2$ and 4.3 A/mg$_{Pt}$, which were 33.9 and 30.7 times higher than commercial Pt/C, respectively. Besides, the highly ordered inner structure of PtPb cores together with the relatively thick Pt shells endowed this material with excellent stability, resulting in a low 7.7% loss in mass activity after 50000 cycles.

In the oxidation of fuel at the anode, CO is a natural intermediate that binds very strongly to the Pt surface, especially at the bridge and 3-fold hollow sites, leading to a dramatic decrease in catalytic efficiency for Pt-based catalysts.[237,238] As such, CO poisoning is a major problem that needs to be addressed in designing more efficient Pt-based catalysts for the oxidation of fuels. Combining Pt with an oxyphilic metal to form bimetallic nanocrystals with an atomic disordered or ordered arrangement provides an effective strategy to mitigate the CO poisoning effect according to the so-called bifunctional mechanism.[239,240] In addition, different from random alloys, ordered intermetallic nanocrystals offer predictable control over structural, geometric, electronic, and ensemble effects. The last two decades have witnessed the successful synthesis of Pt-based intermetallic nanocrystals through the incorporation of oxyphilic metals such as Ti, V, Nb, Ta, Cr, Mn, Fe, Co, Ni, Cu, Ag, Zn, In, Sn, Pb, Sb, and Bi, toward the oxidation of fuel with enhanced catalytic properties.[241-250]

In terms of methanol oxidation (MOR), intermetallic PtPb nanocrystals are promising electrocatalysts in terms of higher activity and superior CO tolerance relative to the commercial Pt/C or PtRu/C. In one example, PtPb intermetallic nanorods with a B8$_1$ structure exhibited a peak current density of over 700 mA/mg in terms of Pt mass for MOR relative to a value of only 450 mA/mg measured from commercial PtRu/C.[201] Further study indicated that the dramatic enhancement of CO tolerance on PtPb intermetallic nanorods could be attributed to the increased Pt-Pt distance, which effectively inhibits CO adsorption at the bridge and 3-fold hollow sites.[251] In another study, Xie and coworkers compared the catalytic activity and CO tolerance of $Pt_3Zn$ intermetallic concave cubes with $Pt_3Zn$ alloy nanocrystals to commercial Pt/C toward MOR.[210] Their results indicated that the intermetallic concave cubes displayed highest forward peak current density (2.58 mA/cm$^2$) and enhanced CO tolerance (higher forward/backward current density ratio) in comparison with the other two catalysts. These results were attributed to the optimal geometric electronic structure and high-index facets.

Ethanol oxidation (EOR) involves C-C bond cleavage to generate a large number of CO intermediates. By incorporating a more electropositive and oxyphilic metal such as Nb and Ta into Pt-based intermetallic nanocrystals, complete electrooxidation of ethanol is promoted since the second metal can more favorably generate additional OH species from water.[252] In a recent study, Abe and coworkers evaluated the EOR properties of $Pt_3Ta$ intermetallic nanocrystals relative to the commercial Pt/C and $Pt_3Sn$/C, showing much lower onset potential, higher current density, and longer cycling life for the intermetallic catalysts.[253] Both CO stripping tests and DFT calculations showed that the substantially enhanced EOR properties could be attributed to the significantly weakened CO chemisorption on the $Pt_3Ta$ intermetallic surface and the strengthened structure stability arising from the low formation enthalpy.

The oxidation of formic acid generally proceeds *via* a dual-pathway mechanism including the direct formation of $CO_2$ through dehydrogenation and an indirect pathway with CO as a reaction intermediate through dehydration.[254] The indirect pathway usually occurs on the monometallic Pt electrocatalysts, leading to significant CO poisoning. Implementing bifunctional intermetallic catalysts is a conventional way to alleviate CO poisoning by promoting the CO oxidation reaction with the oxygenated species. For example, $Pt_3Sn$ intermetallic cubes with three types of surfaces including regular, concave, and defect-rich structures were synthesized and tested as the electrocatalysts for formic acid oxidation reaction (FAOR) in comparison with commercial Pt/C, showing the enhanced oxidation current density for three $Pt_3Sn$ catalysts according to the bifunctional mechanism.[217] Since the defects could facilitate FAOR effectively, the $Pt_3Sn$ nanocubes with a defect-rich surface displayed the best catalytic properties in terms of activity and stability. However, the large overpotential (> 0.6 V *vs.* RHE) is inevitably needed for the oxidation reaction *via* the indirect pathway. To solve this problem, the use of Pt-based intermetallics should lead the oxidation of formic acid through the direct pathway. A recent study showed that the interruption of continuous Pt sites by distinct metal atoms can inhibit the dehydration reaction of FAOR (*i.e.*, ensemble effect) since this indirect pathway requires a continuous array of adjacent Pt sites.[255] To this end, intermetallic structures with highly isolated surface Pt atoms, such as $B8_1$ PtPb and PtBi, were investigated as promising catalysts for FAOR.[195,241,256] Among the earliest examples was that from Abruña and DiSalvo as well as the coworkers, who demonstrated the synthesis of intermetallic PtPb and PtBi nanocrystals with several approaches.[188,194,212,257] According to the results of electrochemical measurements, the catalytic performance of these materials dramatically surpassed the counterparts of catalysts (*e.g.*, Pt black and PtRu nanoparticles) from the aspect of onset potential and current density. The authors also investigated the electrocatalytic kinetics of these intermetallic catalysts with electrochemical mass spectrometry and Fourier transform infrared spectroscopy and thus confirmed that direct pathway took the lead in FAOR catalyzed by intermetallic PtPb and PtBi.[258,259] In another example, Murray and coworkers demonstrated that the dehydration pathway in FAOR could be largely inhibited by using $Pt_3Pb$ intermetallic nanocrystals as the electrocatalysts, leading to nearly a 15-fold enhancement in mass activity at 0.3 V *vs.* RHE, relative to commercial Pt/C.[260] Coating a Pt shell with a thickness of ~0.15 nm on the $Pt_3Pb$ intermetallic nanocrystals promoted the adsorption of formic acid on the catalysts without altering the reaction pathway, resulting in additional 1.7-fold enhancement in mass activity for FAOR. Most recently, Yang and coworkers reported the synthesis of PtAg intermetallic nanocrystals with a mixed hexagonal and cubic closely packed phase by thermal conversion of A1 alloy nanocrystals.[261] The intermetallic phase was only generated in a narrow stoichiometry with Pt/Ag atomic ratio around 47/53 to 48/52 owing to their immiscibility in bulk under equilibrium conditions. Due to the ensemble effect associated with a unique arrangement of Pt and Ag, the PtAg intermetallic nanocrystals were highly active for FAOR at low anodic potentials *via* the direct pathway, showing activities of 5- and 29-fold higher, respectively, as compared to the corresponding alloys and commercial Pt/C at 0.4 V *vs*. RHE in specific activity (Figure 19). Interestingly, Sun and coworkers found that Au could promote the ordering of PtFe in a trimetallic PtFeAu random alloy nanocrystals when thermally annealed in a Ar/$H_2$ gas.[129] The formation of an intermetallic structure was enhanced from the Au atoms segregating to the particle's surface, which left behind lattice vacancies that ultimately enhanced diffusion. The $L1_0$-FePtAu intermetallic nanocrystals exhibited a mass activity of 2809.9 mA/$mg_{Pt}$ at an anodic peak potential and retained 92.5% of this activity after a 13 h stability test toward FAOR, giving the most active and durable catalyst ever reported. The authors demonstrated that the segregated Au shell on the $L1_0$-FePtAu intermetallic nanocrystals could boost the formation of formate in the first step, promoting the oxidation of formic acid through the direct pathway.

## 4.2. Pt-free noble metal intermetallic nanocrystals for catalysis

Due to the scarcity and high cost of Pt, there is an ongoing effort to identify metals with similar catalytic performance but have a higher abundance. One potential candidate is Pd, which has a similar electronic structure and lattice parameters to those of Pt.[262-264] However, the poor durability of Pd-based catalysts in strongly acidic environments presents a major obstacle for their widespread use in fuel-cell technologies. However, the high structural stability typically associated with intermetallic compounds offers a potential route toward extending the catalytic lifetime of Pd-based catalysts while under such corrosive conditions.[265] To this end, Sun and coworkers demonstrated the synthesis of $L1_0$ PdFe intermetallic nanocrystals with different degrees of atomic ordering by thermally annealing Pd@Fe$_2$O$_3$ nanocrystals at various temperatures and for different periods of time. They showed a clear positive trend between the ORR activity and the degree of atomic ordering, as shown in Figure 20a.[266] To further enhance their catalytic performance, thin, continuous Pd shells were produced around the PdFe intermetallic nanocrystals by etching the sample with acetic acid and then carrying out a thermal treatment. Such intermetallic core-shell nanocrystals with a Pd shell of 0.65 nm thick displayed a comparable half-wave potential and mass activity (105 mA/mg$_{Pd}$) relative to the commercial Pt/C (112 mA/mg$_{Pt}$) in acidic media, due to a favorable lattice compression in the Pd shell. During 3000 cycles, the core-shell catalysts showed good stability similar to that of the Pt/C, but were better than L1$_0$ PdFe nanocrystals, which could be attributed to the stabilizing effect of the Pd shell on the intermetallic core. Recently, Pd-based intermetallic core-shell structures with an Au-modified surface was reported to exhibit Pt-like ORR catalytic performance.[165] Favored by their intermetallic structures enclosed with stable facets, the AuPdCo core-shell nanocrystals exhibited a significantly enhanced catalytic activity relative to the Co@AuPd core-shell nanocrystals toward ORR in both acidic and basic solutions. Notably, the intermetallic catalysts showed no change in half-wave potential after 10000 potential cycles in KOH solution, in contrast to the commercial Pt/C, which showed an obvious decay of ~30 mV. The authors argued that the atomic ordering of the intermetallic core and the Au-modified surface both contributed to this boost in stability.

The advancements achieved from implementing Pd-based intermetallic nanocrystals has stimulated an interest in their use in a variety of other energy conversion and storage technologies, such as metal-air batteries. To this end, Goodenough and coworkers evaluated the ORR catalytic properties of intermetallic Pd$_3$Fe nanocrystals in a Li-air battery.[267] Under alkaline conditions, the carbon-supported intermetallic Pd$_3$Fe nanocrystals showed a remarkable enhancement in mass activity relative to the disordered Pd$_3$Fe catalysts, commercial Pd/C, and Pt/C. Such enhancement may have resulted from the unique bond lengths and highly uniform surface structure. Moreover, the ordered $Pd_3Fe$ catalysts exhibited a superior durability relative to the other three catalysts as the current density only dropped by ~2% after 10000 s at 0.8 V *vs*. RHE in comparison with a ~11% loss for the commercial Pt/C. This attractive feature was also presented in the long-term charge-discharge test in a Li-air battery with the ordered $Pd_3Fe$ catalysts only showing a 2.2% increase in round-trip overpotential after 220 cycles, setting the record for the best cycling performance ever reported. The outstanding ORR durability could be attributed to the strong electronic coupling and chemical inertness to Fe-leaching from spatially isolated sites. In another example, DiSalvo and coworkers reported the synthesis of ordered $Pd_3Pb$ nanocrystals as an advanced ORR catalyst under alkaline conditions, showing 4-fold increase in mass activity as compared to commercial Pt/C catalysts after 5000 cycles of ADT (Figure 20b).[268] Interestingly, the intermetallic catalysts showed improved tolerance against methanol during the chronoamperometry measurement in a $O_2$-saturated mixture containing 0.1 M KOH and 0.5 M methanol. After 2 h, the mass activity dropped by ~70%, which is smaller than commercial Pt/C, which dropped by 90%. When applied to Zn-air batteries, the intermetallic catalysts displayed a better performance in terms of cell efficiency as compared to commercial Pt/C.

In addition to enhancing the ORR occurring at the cathode, Pd-based intermetallic nanocrystals are also regarded as promising catalysts for the oxidation of fuels at the anode (*e.g.*, FAOR and EOR).[269] Previous studies indicated that Pd-based catalysts are particularly attractive as electrocatalysts for FAOR due to the direct oxidation pathway, in which the production of CO, the major poisoning species, is inhibited.[270] However, Pd-based catalysts often suffer in terms of long-term stability toward FAOR since intermediate species can poison the surface and also due to the fact that Pd is susceptible to dissolution at high potentials. In a recent study, porous $Pd_3Sn_2$ intermetallic nanobranches were found to exhibit improved mass activity (553.7 $mA/mg_{metal}$) with an enhancement factor of 1.9 and 5.6 relative to the $Pd_3Sn_2$ alloy nanocrystals and commercial Pd black, respectively.[271] Besides the enhancements derived from both geometry and ligand effects, the defect-rich surface (low-coordination atoms) also contributed to the enhancement in catalytic activity. In addition, after 300 cycles, the $Pd_3Sn_2$ intermetallic catalysts retained ~53% of the mass activity as compared to the Pd black, which lost ~95% of the original activity.

Apart from FAOR, Pd-based intermetallic nanocrystals have also been widely explored as electrocatalysts for EOR, which requires high resistance to poisoning as well. For example, Peter and coworkers demonstrated the solvothermal synthesis of hexagonal $Pd_2Ge$ intermetallic nanocrystals with different degrees of crystallinity in order to compare the respective activities and stabilities toward EOR.[272] According to theoretical calculations, the ordered atomic arrangement and Ge vacancies on the surface remarkably changed the absorption energies of the intermediates (*e.g.*, $CH_3CO$ and OH species), which helped prevent poisoning.

Besides the applications in electrocatalysis, Pd-based and other Pt-free intermetallic catalysts also play a key role in facilitating a variety of organic heterocatalytic reactions such as hydrogenation, dehydrogenation, oxidization and steam reforming, where these metal catalysts are usually supported on oxides.[222] In these reactions, intermetallic compounds are generally superior in terms of selectivity and stability relative to their alloy counterparts. As a model example, ZnO supported Pd-Zn nanoparticles with intermetallic structures have proven to be a promising material for the methanol reforming reaction, which is an important approach to the mass-production of $H_2$.[273] Compared to the reductive atmosphere in hydrogenation reactions, better stability of catalysts is a prerequisite in the steam reforming reactions due to the more severe conditions. In an oxidative environment, ultrathin oxide patches or shells are formed on the surface of the intermetallic catalysts, serving as a passivation layer that block further oxidation without destroying the intermetallic structure. Furthermore, through electron microscopic characterization and methanol reforming measurement, Armbrüster and co-workers demonstrated that the ZnO patches on the PdZn intermetallic nanocrystals improved the activation of water and thus the selectivity toward $CO_2$.[274] At the same time, the heteroatomic structures of the tetrahedral interstices in the intermetallics display a strong repulsion to hydrogen attack and dissolution, which contributes to the improved stability in $H_2$ atmosphere as well.[275]

In addition, there are some reports on AuCu-based intermetallic catalysts for electrochemical reduction of $CO_2$ and vapor oxidation of CO, but no favorable result was achieved relative to their alloy phases.[175, 276, 277]

**4.3. Non-noble metal intermetallic nanocrystals for catalysis**

Noble metals are among the rarest elements in the earth's crust. Their costs continue to rise as a result of their unbalanced supply and demand. There is an urgent need to replace the noble metals in catalysts with the less expensive and more abundant base metals. Instead of carrying out trial-and-error studies to screen a daunting number of bimetallic pairs, we can instead turn to theory for assistance. One good example can be found in the semi-hydrogenation of alkynes, which typically utilizes Pd-based catalysts.[278-280] In order to identify a bimetallic alternative, Nørskov and coworkers developed a DFT-based model to screen different catalysts for the semi-hydrogenation of acetylene.[281] In this model, adsorption energies were regarded as descriptors of the activation barriers at key reaction steps and could therefore be employed to determine both the catalytic activity and selectivity. On the basis of this model, several other bimetallic systems, including Co-Ga, Ni-Ga, Fe-Zn, and Ni-Zn, have been identified as both highly active and selective catalysts toward the semi-hydrogenation of acetylene. Among these promising candidates, the B2 Ni-Zn intermetallic catalysts stood out due to their markedly low cost. The theoretical work suggested that the boost in catalytic performance could be primarily attributed to the unique electron structure of the surface, with the Ni atoms serving as the active sites in the reaction. This theoretical prediction was verified through the hydrogenation experiments involving Ni-Zn bimetallic catalysts with different compositions relative to the industrially used Pd and Pd-Ag catalysts (Figure 21). It was found that both NiZn and $NiZn_3$ exhibited favorable performance, and remarkably, NiZn$_3$ was found to have a selectivity close to 100%. In addition to activity and selectivity, the NiZn intermetallic catalysts were expected to show excellent stability due to the relatively high segregation energy for both components.

Inspired by the aforementioned work, Wang and coworkers recently demonstrated that Ni$_x$M$_y$ (M = Ga and Sn) intermetallic nanocrystals were highly active and selective as catalysts for the semi-hydrogenation of both liquid-alkynes and gas-alkynes.[200] For the hydrogenation of phenylacetylene, all of the Ni-based catalysts showed a much better selectivity and durability as compared to the benchmarks such as Pd/C and Lindlar catalysts. In addition, the Ni$_3$Ga and Ni$_3$Sn intermetallic catalysts showed a conversion as high as 99%. In the semi-hydrogenation of acetylene, Ni$_3$Ga, Ni$_5$Ga$_3$, and Ni$_3$Sn$_2$ intermetallic nanocrystals displayed a selectivity of 80%, which was much higher than supported Pd-Ag catalysts with a selectivity of 55%. The enhancement in selectivity and activity was attributed to the modified electronic structure of the Ni atoms arising from their covalent bonding with Ga or Sn, together with the partial isolation of Ni active sites that are surrounded by the heteroatoms in the ordered atomic arrangement. In addition, the Ni atoms in the intermetallic nanocrystals displayed enhanced resistance against surface oxidation upon adding the second metal. This work demonstrates that intermetallic catalysts could be designed according to the principles of site-isolation and surface electronic modification. It also sheds light on non-noble metal catalysts configured with p-d hybridization as an alternative to the conventional d-d hybridized noble metal based catalysts.

It is also noteworthy to mention that non-noble metal intermetallic catalysts have also been used as substitutes in catalytic reactions that demand high efficiency and low cost. For example, intermetallic Ni-In nanocrystals with tailored electronic structures could be used for preferential hydrogenation favoring nucleophilic addition rather than electrophilic addition, which could selectively hydrogenate unsaturated aldehydes to alcohols.[282] In another study, intermetallic NiGa catalysts with engineered compositions and structures were found to have a suitable adsorption energy for oxygen, and thus exhibited interesting catalytic performance toward the reduction of $CO_2$ to methanol with remarkably lower production of CO.[283]

## 5. Concluding remarks and outlook

Intermetallic nanocrystals represent an exciting and vibrant subject of research converged at the forefronts of materials chemistry, catalysis, and nanotechnology. Compared to their alloy counterparts, intermetallic nanocrystals often exhibit enhanced catalytic properties in terms of activity, selectivity, and durability for a variety of reactions. The superb catalytic properties can be attained by precisely engineering their composition, size, and shape, which in turn directly impact the surface atomic structure. Relative to the significant progress made for bimetallic nanocrystals with an alloy structure, the production of intermetallic nanocrystals, especially for those with well-controlled shapes, is still in the early stages of development due to the lingering challenges and constraints imposed by the complex thermodynamic and kinetic parameters involved during a synthesis. A number of scientific problems or unsolved issues still need to be addressed. One of the major challenges is to simultaneously achieve control over both the intermetallic composition and the shape. Although thermal conversion is effective in generating intermetallic structures from alloys, annealing at elevated temperatures inevitably deteriorates the shape of the original templates in most cases. In contrast, wet chemical approaches hold great promise for the generation of intermetallic nanocrystals with well-defined shapes by employing a mild reducing agent at a relatively low temperature to controllably slow down the reaction kinetics. In particular, the one-pot approach would offer a simple and viable approach to intermetallic nanocrystals with well-defined shapes. However, the potential of this method tends to be masked by the highly complex and enigmatic homogenous nucleation process, which is still a black box even for the monometallic system.

Among the various methods, seed-mediated diffusion growth in a solution phase has been most productive in generating intermetallic nanocrystals with controlled shapes. Implementing seeds that already have an alloy or intermetallic composition increases the likelihood for interdiffusion, and thus favors the formation of intermetallics. In addition, the internal structure of the seeds (*e.g.*, single-crystal *vs.* multiply-twinned) can also be explored to engineer intermetallic nanocrystals. In principle, the well-developed ideas associated with both thermodynamic and kinetic controls over the monometallic system can be similarly employed for shape-controlled syntheses of intermetallic nanocrystals in the setting of seed-mediated growth. For example, the use of capping agents can still serve as a general means to maneuver the shape of intermetallic nanocrystals through selective adsorption on a specific type of facet. However, it should be noted that the adsorption behavior for the capping agent used for a monometallic system should be different from the intermetallic varieties, especially for those with a structural evolution during chemical conversion (*e.g.*, from A1 to $L1_0$). To be more efficient in identifying effective capping agents, efforts should be devoted to better understand the adsorption behavior of capping agents on intermetallic nanocrystals by theoretical calculations. As another note, it should be pointed out that kinetic control has been established as a powerful means for generating metallic nanocrystals with a concave or core-shell structure. Achieving the same level of control can be expected for intermetallic nanocrystals, but it will require some new developments in terms of synthetic strategies.

Thermodynamic and kinetic analyses show that the formation of intermetallic nanocrystals is regulated by a number of factors, including the surface free energy, activation barrier, and diffusion barrier. All these factors play a formative role in dictating the final size, shape, composition, and structure of the nanocrystals. These fundamental parameters can serve as essential guidelines for generating intermetallic nanocrystals with desired features and thus properties. We have discussed a large number of recently reported strategies based on thermal conversion from a random alloy or separated components, seed-mediated diffusion growth, and one-pot synthesis. Based on the discussions, we have summarized the synthetic routes that facilitate the formation of intermetallic structures with controlled shapes. The comparison of intermetallic structures with random alloy phases used as catalysts in some typical catalytic reactions not only show their fascinating catalytic properties but also allow us to correlate size, shape, composition, and structure with function and performance, providing key design criteria for developing advanced intermetallic catalysts.

In order to address the cost issue in industrial applications, partially or completely replacing noble metals with non-noble metals without compromising the catalytic properties is a highly desired but extremely difficult research endeavor. As discussed in Section 4.3, numerous examples of non-noble metal intermetallic catalysts have exhibited comparable or even better catalytic properties relative to the industrial catalysts based on noble metals for a given reaction due to the unique chemical and structural features. However, owing to their low reduction potentials, the synthesis of non-noble metal intermetallic nanocrystals in a solution phase typically requires a strong reducing agent, making it very difficult to control the shape. However, formation through decomposition of their specific metal precursors (*e.g.*, carbonyl compounds) in oil-phase solutions might be a promising method to overcome this problem.

Synthesizing intermetallic nanocrystals that contain more than two elements offers additional tunability in terms of optimizing their catalytic properties for a given reaction (see Section 4.1 for the use of $L1_0$ PtFeCu nanocrystals as ORR catalysts). However, the incorporation of multiple elements will significantly increase the complexity of both the synthesis and mechanisms involved. When a third component has to be incorporated, the most effective strategy should involve further upgrading already effective bimetallic systems. Taking PtFe@Pt nanocrystals as an example (see the discussion in Section 4.1), by replacing some of the Fe atoms with Cu atoms, the larger size associated with Cu atoms could effectively lessen the compressive strain in the Pt-skin layer and thus increase the activity. The similar strategy has also been applied to other highly effective bimetallic catalysts, for example, PtPb@Pt-skin nanoplates.

Furthermore, recent studies suggest that core-shell nanocrystals with ultrathin shells serve as an effective catalytic system with a number of attractive features, including substantially enhanced catalytic properties, remarkably reduced use of rare and expensive metals, and a well-defined surface structure.[284-286] By taking advantage of their high stability in harsh environments and their predictable electronic and lattice structures, intermetallic nanocrystals are expected to serve as better candidates for seeding the formation of core-shell nanocrystals through seed-mediated epitaxial growth. It is also foreseeable that the intermetallic cores could force the ultrathin shell to take on a unique structure that would otherwise be unattainable.

Taken together, research into the shape-controlled synthesis of intermetallic nanocrystals and their utilization as catalysts provides great opportunities that coexist with many challenges. The vast importance of this topic is exemplified by the large number of research groups that have worked arduously to cultivate our understanding of this blossoming field. It is without question that these efforts will pay off in a great way, as the best is yet to come.

**Acknowledgements**

This work was supported in part by grants to H.Z. from the National Science Foundation of China (51372222 and 51522103), National Program of Top-notch Young Professionals (2014), and the Fundamental Research Funds for the Central Universities (2015XZZX004-23); and grants to Y.X. from the National Science Foundation (CHEM 1505441 and DMR 1506018).

Received: ((will be filled in by the editorial staff))
Revised: ((will be filled in by the editorial staff))
Published online: ((will be filled in by the editorial staff))

**Table 1.** Summary of intermetallic nanocrystals with well-defined shapes synthesized by wet-chemistry approach and their reported applications.

| Composition | Crystal structure | Shape | Facet | Application | Ref. |
|---|---|---|---|---|---|
| AuCu | *fct* | pseudo-icosahedra, -decahedra | / | / | 173 |
| AuCu$_3$ | *fcc* | pseudo-icosahedra, -decahedra | / | / | 173 |
| AuCu | *fct* | rod | {100}, {111} | C–N coupling | 174 |
| AuCu$_3$ | *fcc* | rod | {100}, {111} | C–N coupling | 174 |
| Au$_3$Cu | *fcc* | truncated cube | {100} | CO$_2$ reduction | 175 |
| PtPb | *hcp* | rod | {110} | MOR[A] | 201 |
| PtPb@Pt | *hcp@fcc* | plate | {110}Pt//{010}, {001}PtPb | ORR[A], MOR[A], EOR[A] | 206 |
| PtBi | *hcp* | plate | {101} | MOR[A], FAOR[A] | 205 |
| Pt$_3$Zn | *fcc* | concave cube | {520}, {210}, {530} | MOR[A] | 210 |
| Pt$_3$Sn | *fcc* | cube | {100} | / | 208 |
| PtSn | *hcp* | polyhedron, wire | / | / | 208 |
| Pt$_3$Co | *fcc* | nanowire | {310}, {110} | ORR[A] | 209 |
| Pt$_3$Sn | *fcc* | concave cube | {830}, {740}, or {540} | FAOR[A] | 213 |
| PtCu$_3$ | *fcc* | cage | high-index | MOR[A] | 218 |
| PdCu | *bcc* | cube | {100} | ORR[B] | 202 |
| Pd$_2$Sn | *ortho.* | rod | {100}, {001}, {102} | EOR[B] | 203 |
| FeSn$_2$ | *tetr.* | hollow square, U-shape, rod dimer | {100} | / | 179 |
| NiZn | *bcc* | hollow sphere | / | / | 180 |

*Ortho*. = Orthogonal, and *Tetr*. = Tetragonal. [A]Reaction carried out under acidic condition, and [B]reaction carried out under alkaline condition.

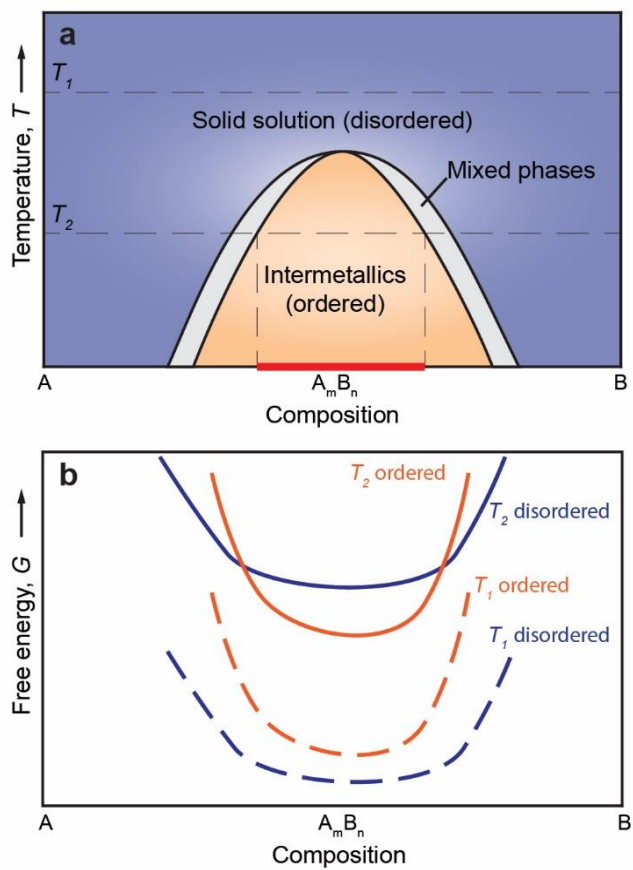

**Figure 1.** Schematic illustration of thermodynamic equilibrium conditions for the formation of an intermetallic phase ($A_mB_n$) in the A-B binary system. a) Phase diagram showing the equilibria among a disordered solid solution (blue), an ordered intermetallic compound (orange), and mixed phases (grey) as a function of composition and temperature. At temperature $T_1$ ($T_1$ is supposed to be higher than the order-to-disorder transition temperature), a disordered atomic arrangement is preferred over the full range of possible compositions. At temperature $T_2$, an ordered phase appears within a specific range of compositions near the stoichiometric point (marked by a red bar). b) Diagram showing the Gibbs free energies of the disordered (blue) and ordered (orange) phases as a function of composition at two different temperatures of $T_1$ (dashed) and $T_2$ (solid), respectively.

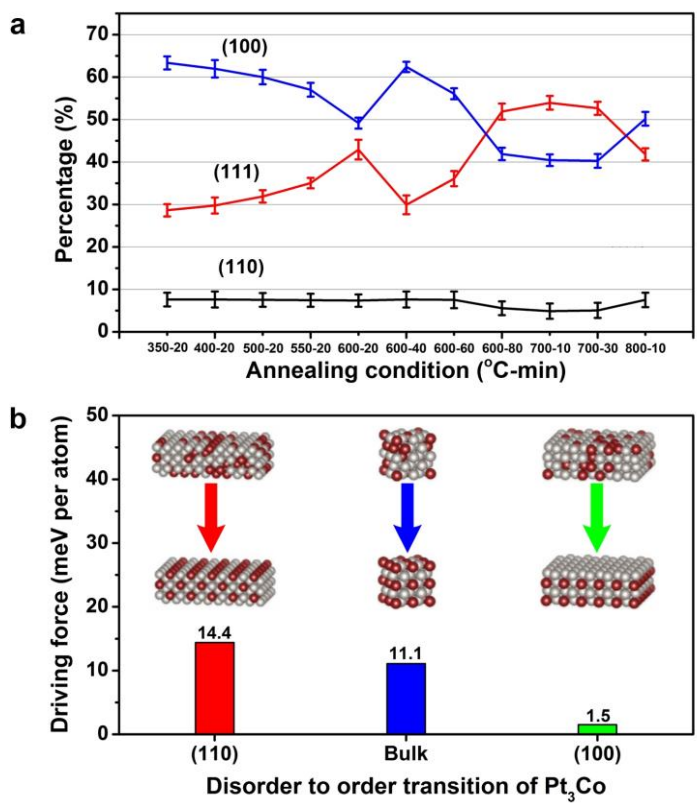

**Figure 2.** The thermodynamic and kinetic behaviors associated with surface faceting during the disorder-to-order transition for Pt$_3$Co nanocrystals. a) Percentages of the areas of {111}, {110} and {100} facets as a function of the annealing temperature and time. b) A comparison of the driving force for generating the Pt$_3$Co intermetallic phase at two different crystallographic facets and in the bulk as derived from DFT calculations. Reprinted with permission from ref [115]. Copyright 2016 Nature Publishing Group.

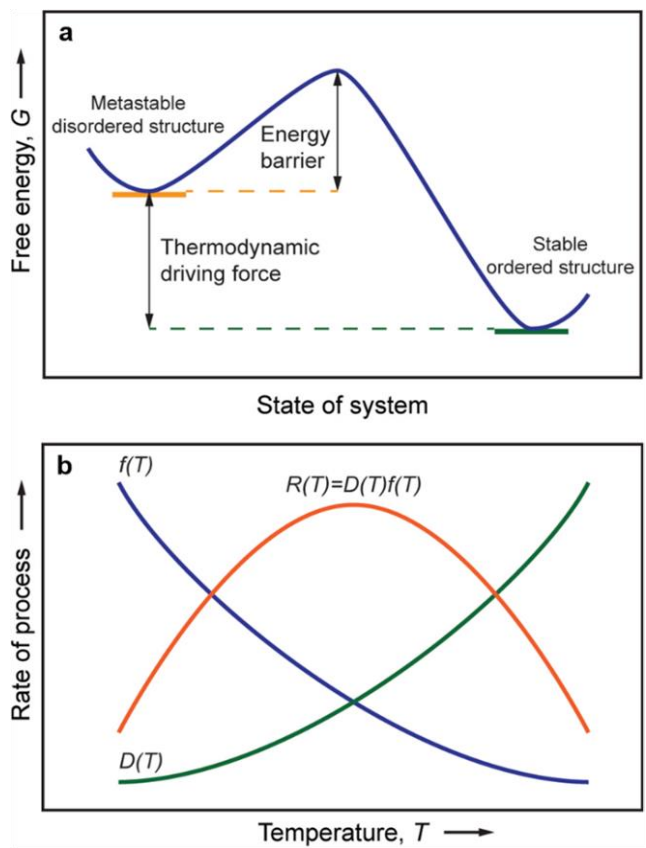

**Figure 3.** a) Schematic illustration of the transition from a metastable disordered phase to a stable ordered phase, in which the transition kinetics is determined by the competition between the energy barrier and the thermodynamic driving force. b) Schematic illustration showing the temperature dependences for the rate of diffusion *D(T)*, rate of new phase formation *f(T)*, and the resulting transition rate *R(T)*, respectively.

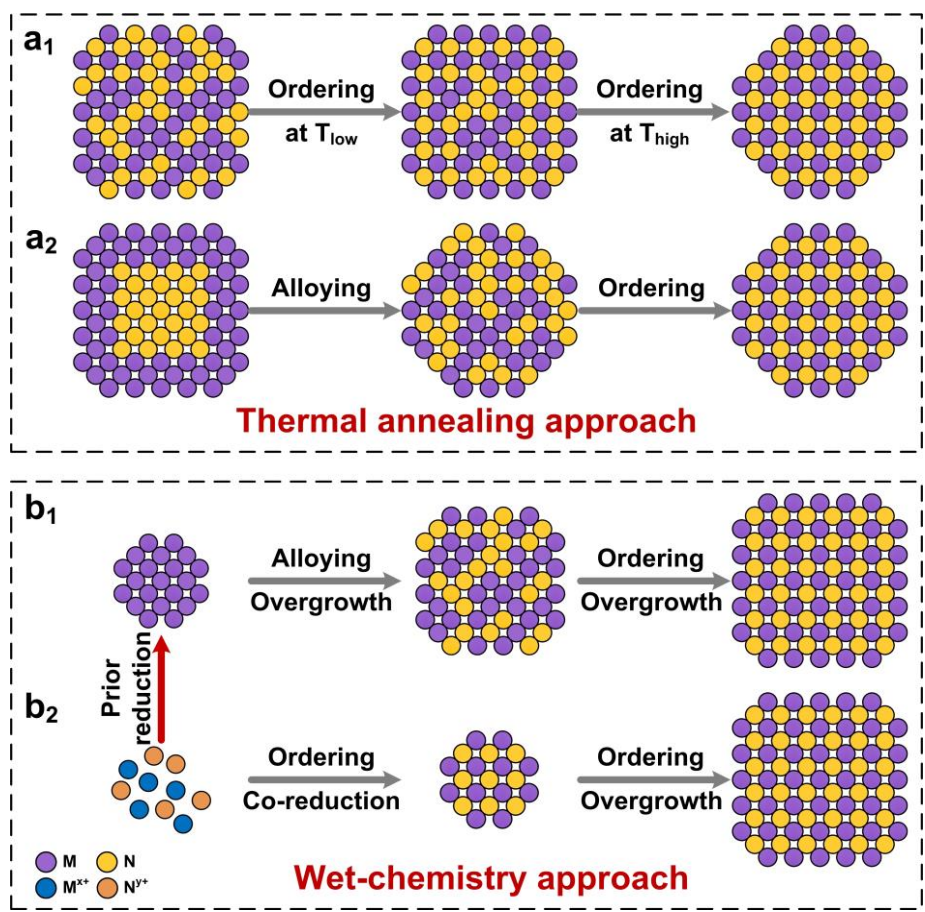

**Figure 4.** Schematic illustration showing the two main strategies used to produce intermetallic nanocrystals: ($a_1$, $a_2$) thermal annealing approach and ($b_1$, $b_2$) wet-chemistry approach. The thermal annealing approach mainly involves the conversion from preformed ($a_1$) alloy and ($a_2$) heterogeneous nanocrystals into intermetallic phases at elevated temperatures in an atmosphere (or vacuum). The wet-chemistry route involves the interplay of reduction, overgrowth, alloying and ordering in solution through ($b_1$) seed-mediated diffusion growth and ($b_2$) one-pot synthesis. The one-pot synthesis can also follow a two-step process similar to seed-mediated diffusion growth, as marked by a red arrow.

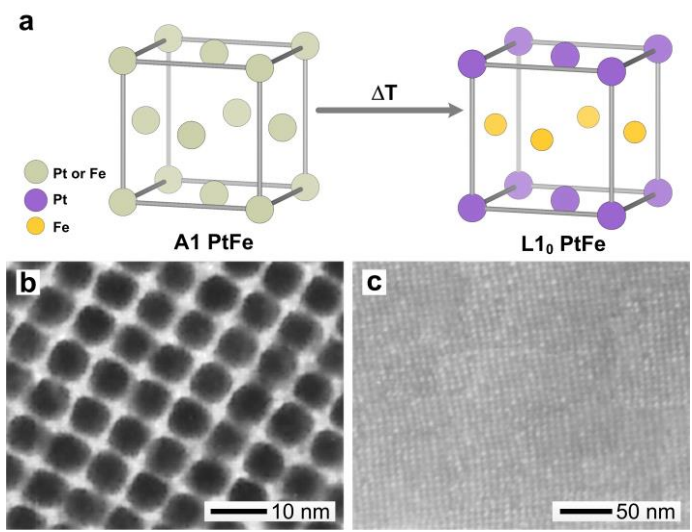

**Figure 5.** a) Schematic illustration of a PtFe unit cell undergoing a phase transition from the A1 alloy phase to the $L1_0$ intermetallic phase under thermal annealing. b) TEM and c) SEM images of the cubic PtFe nanocrystals before and after the thermal annealing treatment, respectively. The images in (b) and (c) were reprinted with permission from ref [146]. Copyright 2000 AAAS.

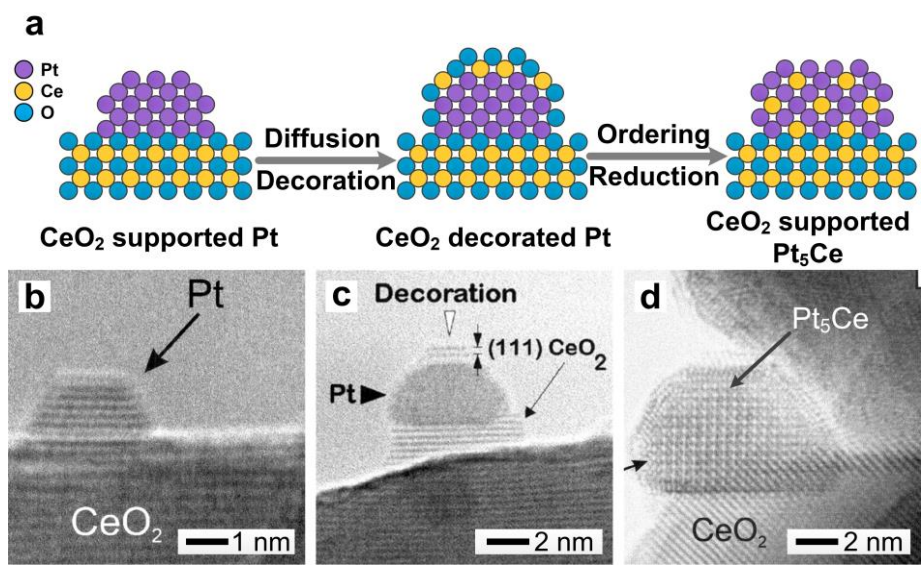

**Figure 6.** a) Schematic illustration showing the formation of CeO$_2$-supported Pt$_5$Ce intermetallic nanocrystals through a reactive metal-support interaction (RMSI) process. The corresponding HRTEM images show b) a Pt nanocrystal supported on CeO$_2$, c) decoration of the Pt nanocrystal by CeO$_2$, d) the Pt$_5$Ce intermetallic nanocrystal after annealing in H$_2$ at 500 ºC. The images in (b–d) were reprinted with permission from ref [169]. Copyright 1997 Elsevier.

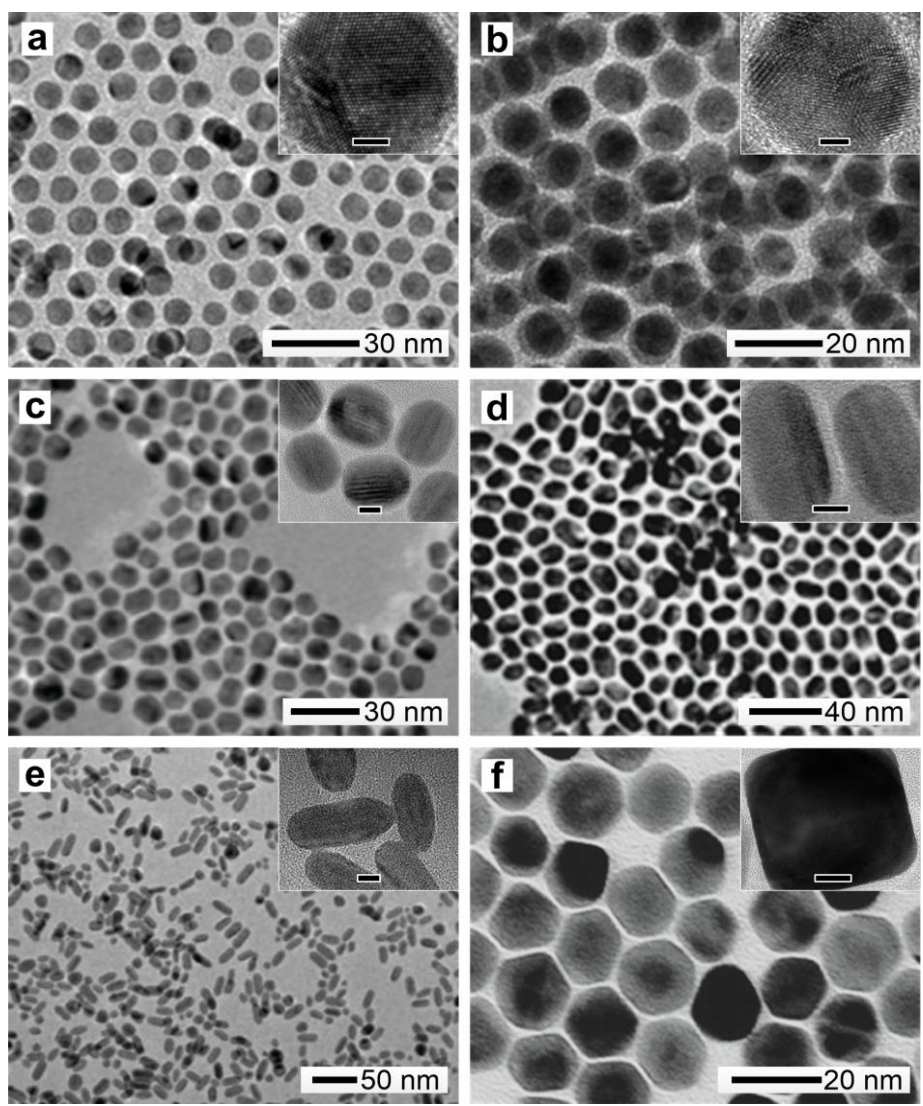

**Figure 7.** a, b) TEM images of multiply-twinned AuCu and AuCu$_3$ pseudo-spherical intermetallic nanocrystals. c–e) TEM images of pseudo-icosahedral or decahedral AuCu, AuCu$_2$, and AuCu$_3$ intermetallic nanorods. f) TEM image of Au$_3$Cu intermetallic cuboctahedra. All insets show the corresponding HRTEM images, with the scale bars corresponding to (a, b) 2 nm, (c–e) 5 nm, and (f) 7 nm. The images in (a, b) were reprinted with permission from ref [173]. Copyright 2010 Wiley-VCH. The images in (c–e) were reprinted with permission from ref [174]. Copyright 2014 The Royal Society of Chemistry. The image in (f) was reprinted with permission from ref [175]. Copyright 2014 The Royal Society of Chemistry.

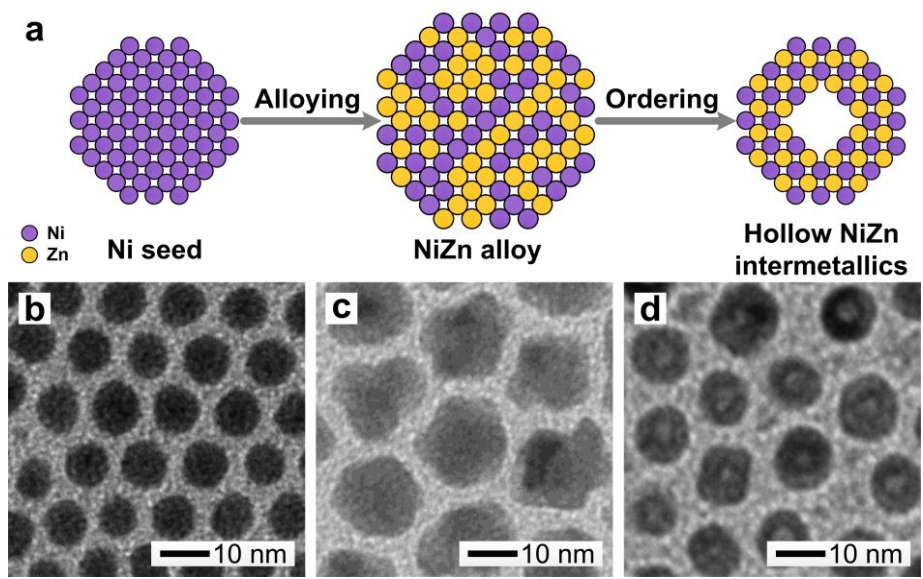

**Figure 8.** a) Schematic illustration showing the formation of hollow NiZn intermetallic nanocrystals derived through the Kirkendall process. The corresponding TEM images show b) the Ni seeds, c) the NiZn alloys at 10 min, and d) the final products at 30 min after the injection of the Zn precursor into the reaction solution. The TEM images were reprinted with permission from ref [180]. Copyright 2013 American Chemical Society.

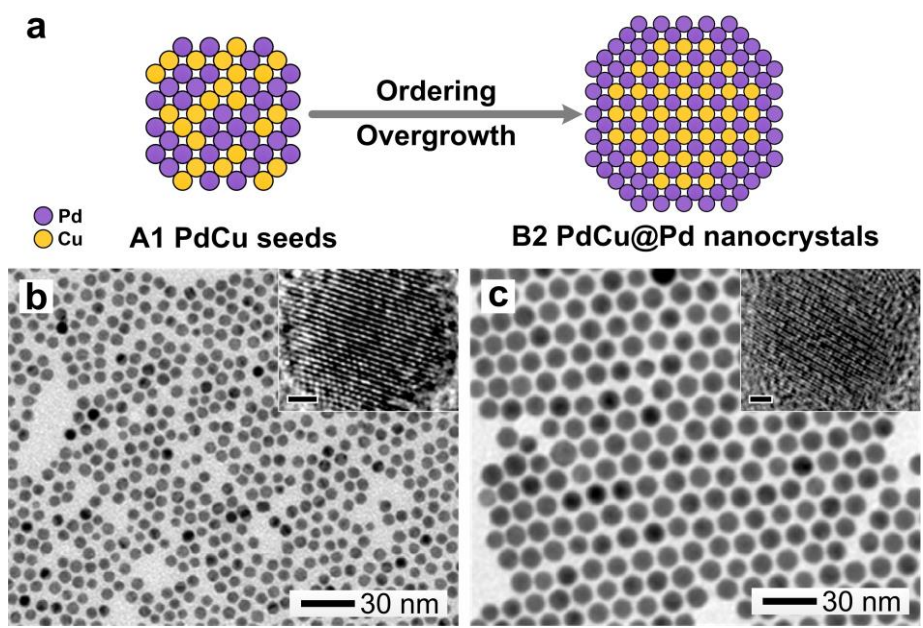

**Figure 9.** a) Schematic illustration showing the formation of B2 PdCu@Pd core-shell nanocrystals with an intermetallic core through seed-mediated diffusion growth in combination with a co-reduction process. TEM images of b) the A1 PdCu seeds with an average size of 6.8 nm and c) the B2 PdCu@Pd core-shell nanocrystals with an average size of 11.9 nm. The insets show the corresponding HRTEM images of individual particles, each with a scale bar of 1 nm. The TEM images were reprinted with permission from ref [133]. Copyright 2016 American Chemical Society.

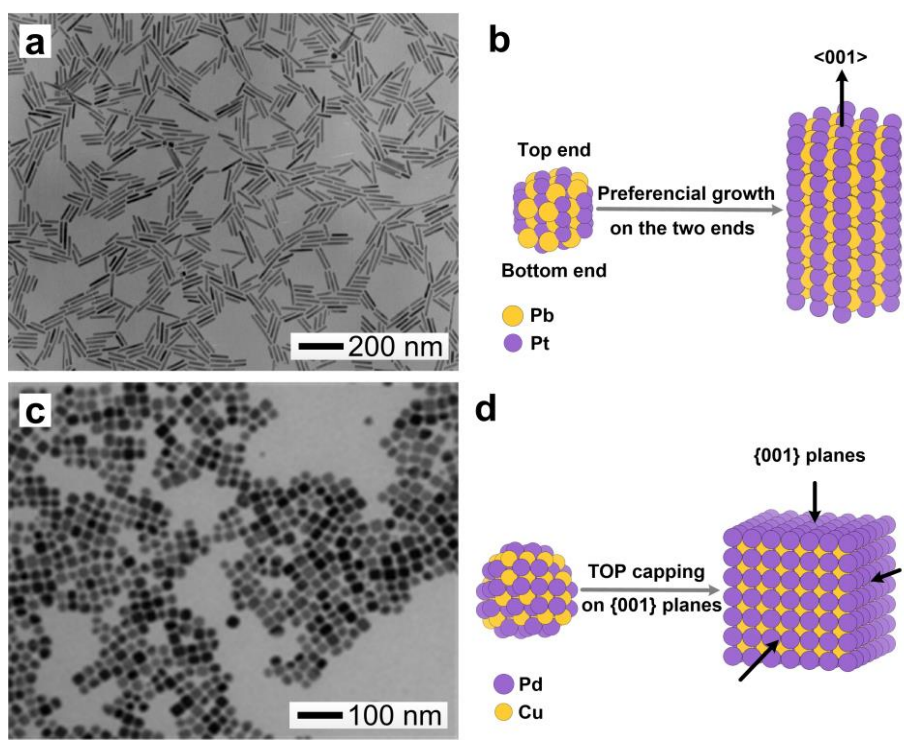

**Figure 10.** a) TEM image of the B8$_1$ PtPb nanorods prepared in a one-pot synthesis. b) Schematic illustration showing the anisotropic growth of the B8$_1$ PtPb nanorods along the <001> direction as driven by the anisotropic hexagonal crystal structure. c) TEM image of the B2 PdCu nanocubes prepared in the presence of TOP as a capping agent. d) Schematic illustration showing the formation of the B2 PdCu nanocubes by selectively capping the {100} planes with TOP. The image in (a) was reprinted with permission from ref [201]. Copyright 2007 American Chemical Society. The image in (c) was reprinted with permission from ref [202]. Copyright 2013 Wiley-VCH.

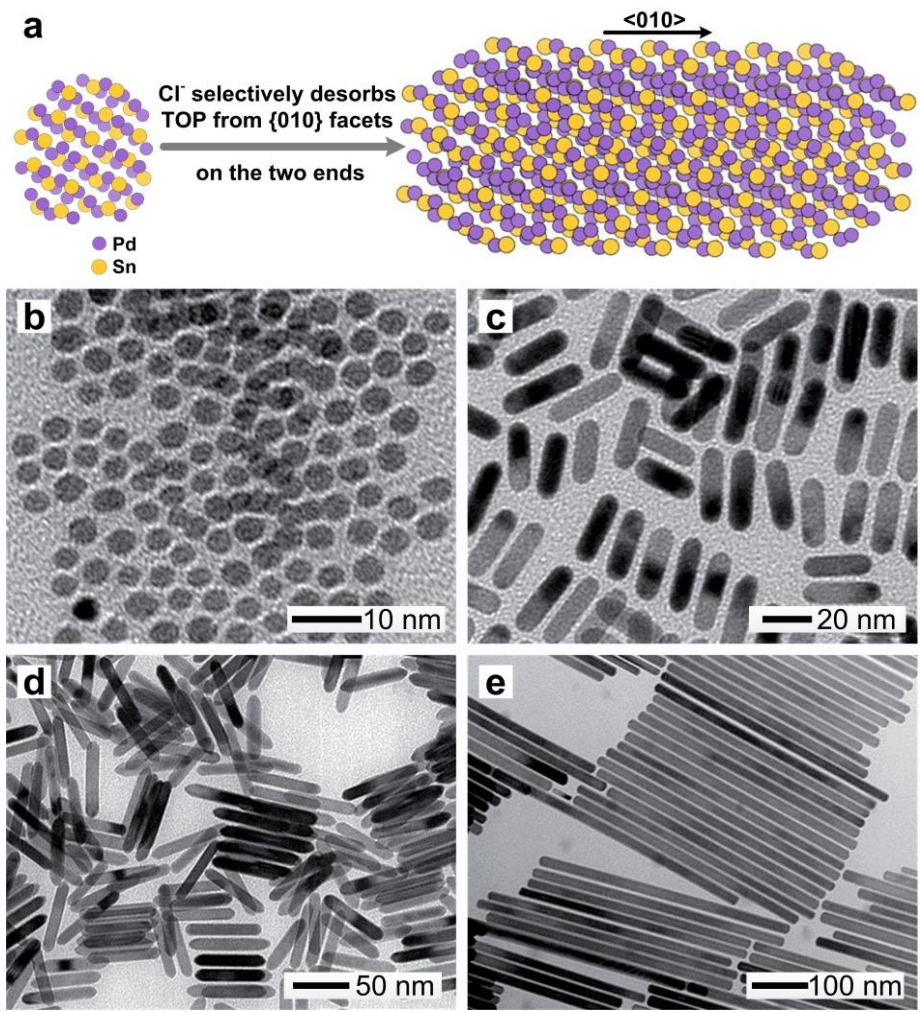

**Figure 11.** a) Schematic illustration showing the anisotropic growth of orthogonal Pd$_2$Sn nanorods along the <010> direction due to the selective desorption of TOP from {010} by Cl$^-$ ions. b–e) TEM images of the Pd$_2$Sn nanoparticles and nanorods with different lengths and diameters achieved by varying the concentration of Cl$^-$ ions and TOP. The images were reprinted with permission from ref [203]. Copyright 2016 The Royal Society of Chemistry.

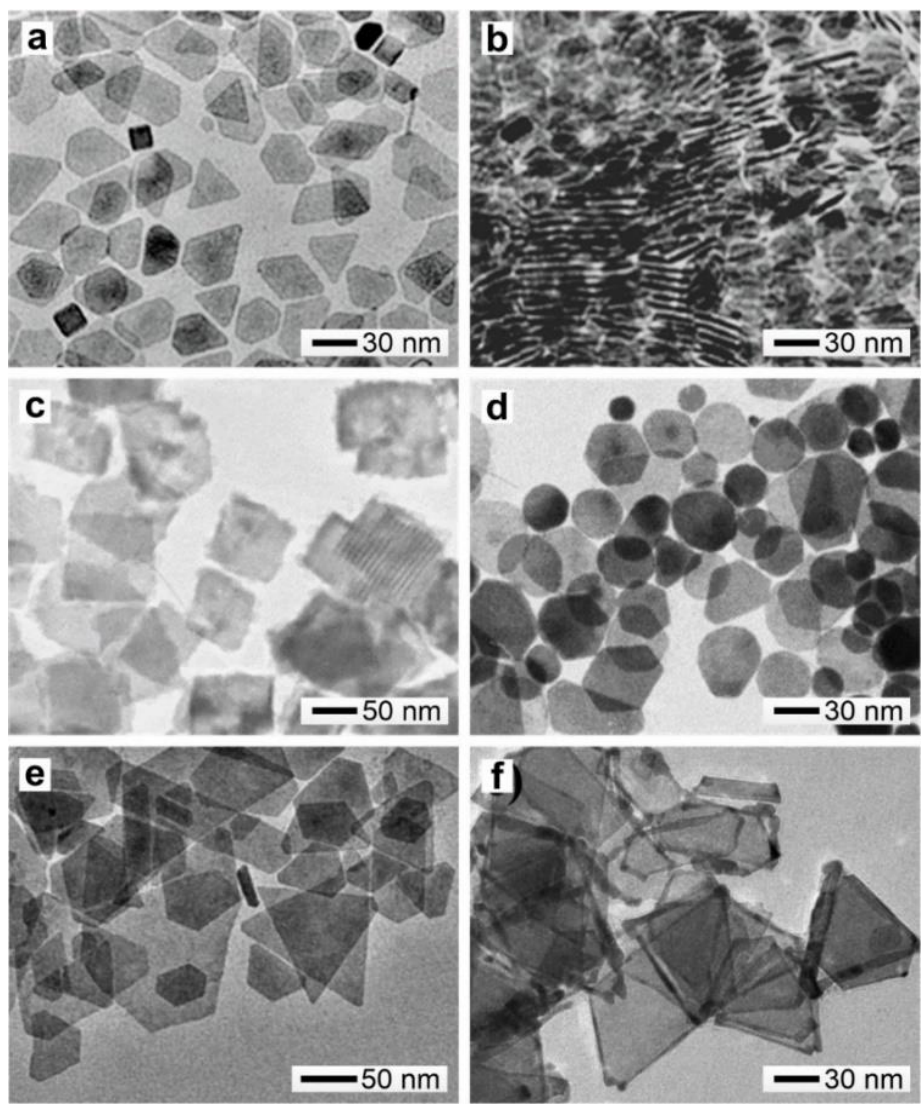

**Figure 12.** a, b) TEM images of PtBi intermetallic nanoplates with a $B8_1$ structure that were prepared by simultaneously reducing $Pt(acac)_2$ and bismuth neodecanoate in OAm at 200 °C with $NH_4Br$ as a capping agent. c, d) TEM images of $B8_1$ PtBi in the form of square nanoplates and nanodisks, which were prepared using bismuth octoate as a precursor to Bi and $H_2PtCl_6$ as a precursor to Pt. e, f) TEM images of $B8_1$ PtBi nanoplates that were prepared with $NH_4Cl$ and $NH_4I$ as the capping agent instead of $NH_4Br$. Reprinted with permission from ref [205]. Copyright 2014 The Royal Society of Chemistry.

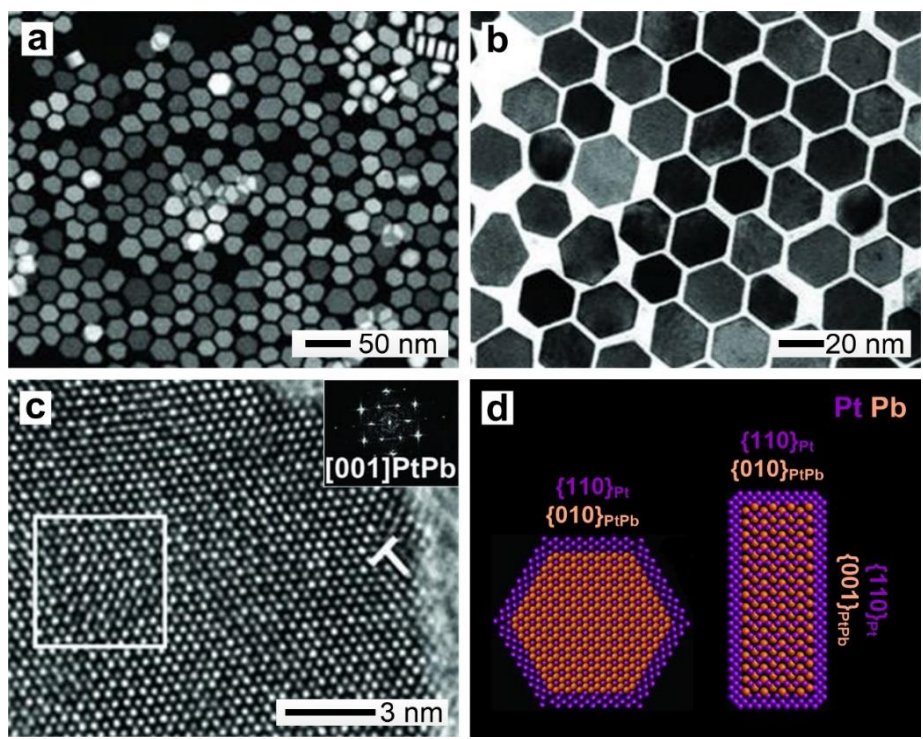

**Figure 13.** a) High-angle annular dark field scanning TEM (HAADF-STEM), b) TEM, and c) atomic-resolution HRTEM images of core-shell PtPb@Pt nanoplates with B81 PtPb cores and A1 Pt shells prepared in a one-pot synthesis. The inset in (c) shows the FFT pattern corresponding to the selected zone. d) Atomic models of the nanoplate displaying the intermetallic and core-shell structure with top interface of [(110)Pt//(100)PtPb] and side interface of [(110)Pt//(001)PtPb]. Reprinted with permission from ref [206]. Copyright 2016 AAAS.

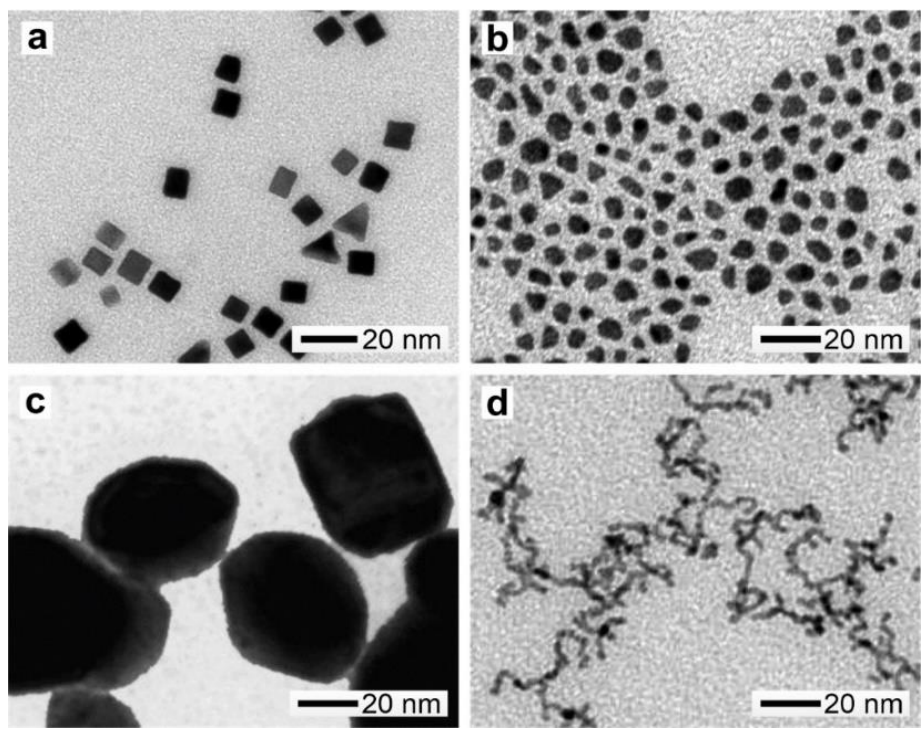

**Figure 14.** a–c) Morphological and structural evolution of dodecylamine (DDA)-capped PtSn intermetallic nanocrystals with different molar ratios of Pt to Sn precursors. TEM images of intermetallic a) $Pt_3Sn$ nanocubes, b) PtSn polyhedra, and c) large irregular $PtSn_2$ particles prepared at Pt/Sn molar ratios of 70/30, 40/60, and 20/80, respectively. d) TEM image of PtSn intermetallic nanowires synthesized by replacing DDA with a small amount of OAm and OA in the presence of 56–74 mol % of Sn. Reprinted with permission from ref [208]. Copyright 2012 American Chemical Society.

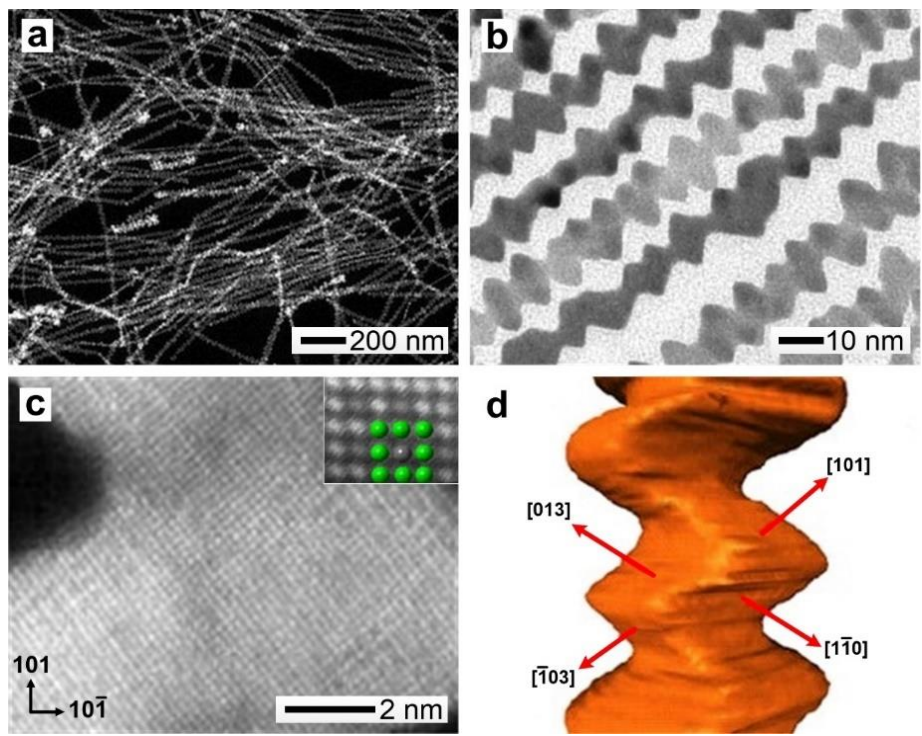

**Figure 15.** a) HAADF-STEM, b) TEM, and c) atomic-resolution HAADF-STEM images of the crenel-like intermetallic Pt$_3$Co hierarchical nanowires with a L1$_2$ structure, high-index facets, and a Pt-rich surface prepared in a one-pot synthesis. d) Schematic illustration showing the high-index {310} facets exposed on the surface of a nanowire. The inset in (c) is an atomic resolution HAADF-STEM image of a Pt$_3$Co nanowire at a higher magnification, showing the L1$_2$ arrangement using green dots to represent Pt atoms and a gray dot to represent the center Co atom. Reprinted with permission from ref [209]. Copyright 2016 Nature Publishing Group.

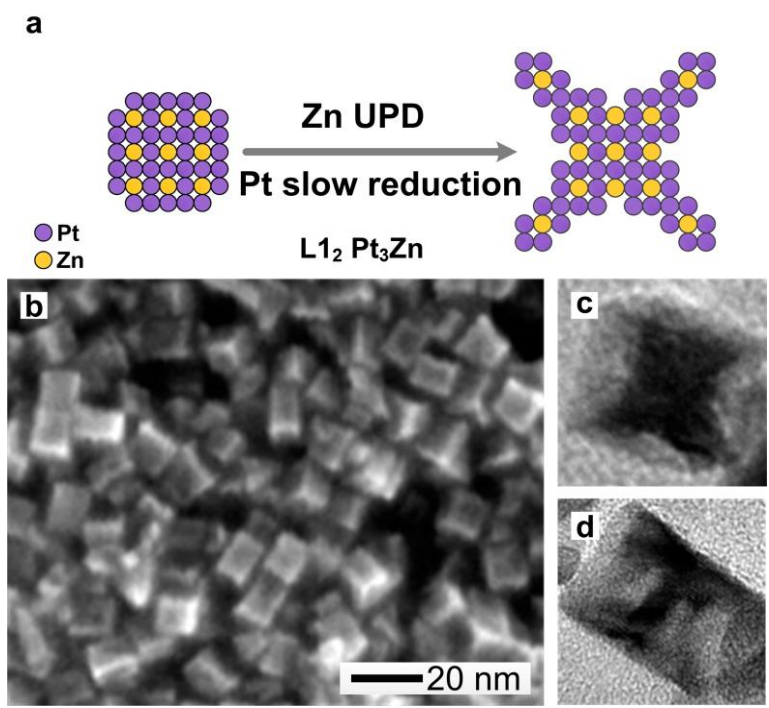

**Figure 16.** a) Schematic illustration showing the formation of the L1$_2$ Pt$_3$Zn concave cubes through a process involving Zn UPD and Pt slow deposition. b) SEM image of L1$_2$ Pt$_3$Zn concave cubes prepared by a solvothermal process in DMF. c, d) TEM images of a single L1$_2$ Pt$_3$Zn concave cube viewed along different angles. The images in (b–d) were reprinted with permission from ref [210]. Copyright 2014 The Royal Society of Chemistry.

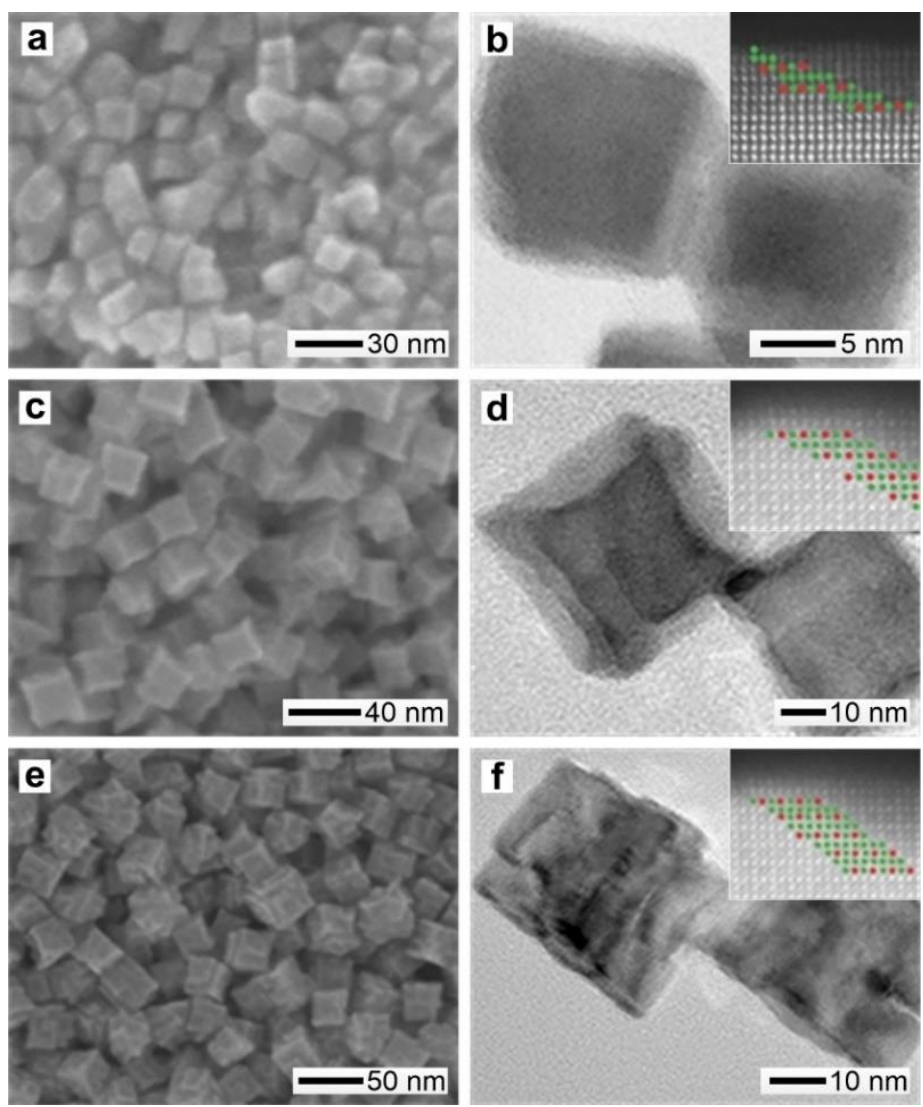

**Figure 17.** SEM and TEM images of a, b) cubic, c, d) concave cubic and e, f) defect-rich cubic intermetallic Pt$_3$Sn nanocrystals prepared using a solvothermal process in DMF under kinetic control by varying the concentrations of the two metal precursors. The insets in (b, d, f) are the corresponding atomic-resolution HAADF-STEM images, where the arrangements of Pt and Sn atoms are marked by green and red dots, respectively. Reprinted with permission from ref [213]. Copyright 2016 Wiley-VCH.

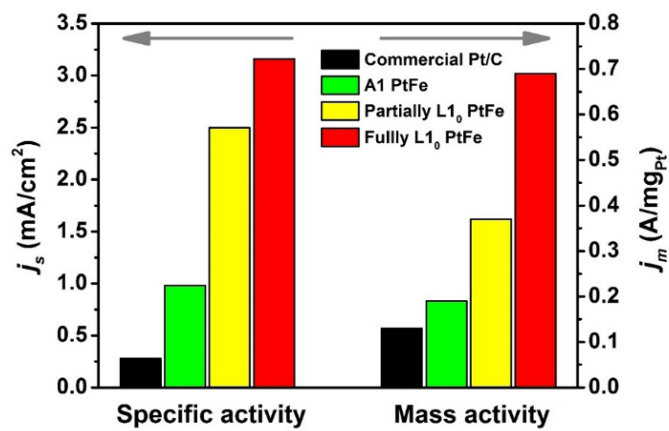

**Figure 18.** Comparison of the specific and mass activities toward ORR for PtFe nanocrystals with three different atomic arrangements including A1 alloy, partially ordered, and fully ordered, in comparison with commercial Pt/C. The data were obtained from the Tafel plots at 0.9 V *vs*. RHE. Reprinted with permission from ref [157]. Copyright 2015 American Chemical Society.

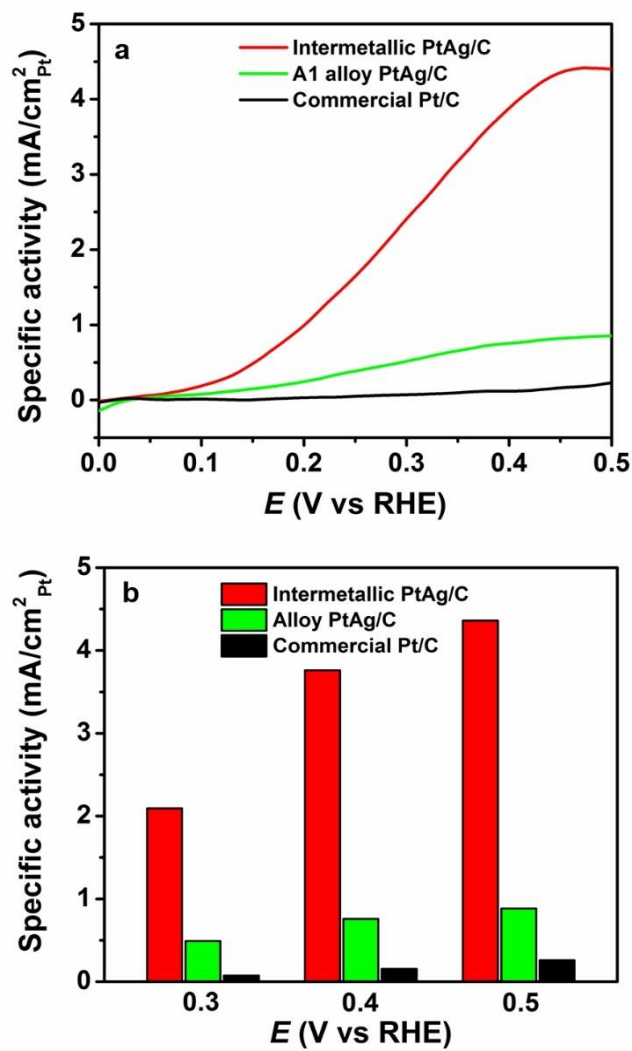

**Figure 19.** Electrocatalytic performance of carbon-supported intermetallic and alloy PtAg nanocrystals towards FAOR, as compared with commercial Pt/C. a) Partial current density-potential curves measured in a mixed solution of 0.5 M formic acid and 0.1 M HClO$_4$. b) Area-specific activity at 0.3, 0.4, and 0.5 V (*vs*. RHE) for intermetallic and alloy PtAg catalysts relative to Pt/C. Reprinted with permission from ref [261]. Copyright 2016 American Chemical Society.

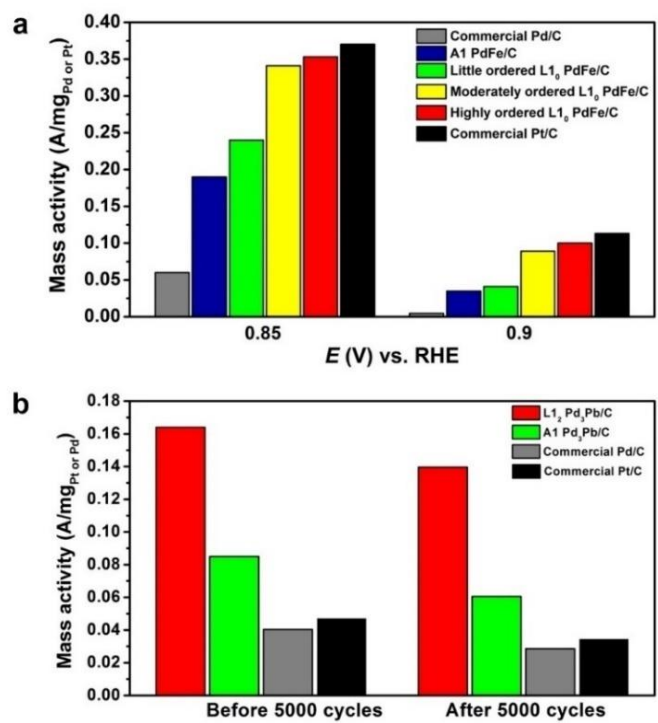

**Figure 20.** Catalytic performance toward ORR for different Pd-based intermetallic nanocrystals relative to the corresponding disordered alloys and commercial catalysts. a) Mass activity of L1$_0$ PdFe intermetallics with different degrees of atomic ordering measured in a 0.1 M HClO$_4$ solution compared to commercial Pd/C and Pt/C. b) Mass activity of L1$_2$ and A1 Pd$_3$Pb nanocrystals measured in 0.1 M KOH solution compared to commercial Pd/C and Pt/C. The catalysts were also compared after 5000 cycles of accelerated durability test. The bar plot in (a) was reprinted with permission from [266]. Copyright 2014 American Chemical Society. The bar plot in (b) was reprinted with permission from ref [268]. Copyright 2016 American Chemical Society.

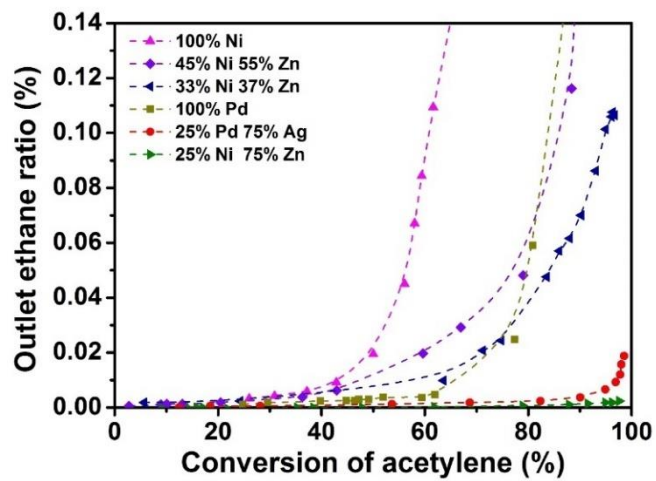

**Figure 21.** Comparison of the selectivity for six different catalysts toward the semi-hydrogenation of acetylene, which was measured with a mixed inlet gas containing 1.33% ethylene, 0.0667% acetylene, and 0.67% $H_2$ at 1 bar. As a function of the conversion of acetylene, the measured concentration of ethane at the reactor outlet was used as an indicator of selectivity, where less production of ethane corresponds to higher selectivity of the catalyst. Reprinted with permission from ref [281]. Copyright 2008 AAAS.

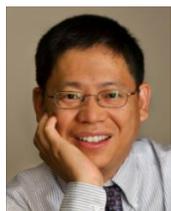

**Younan Xia** received his Ph.D. from Harvard University in 1996 with George M. Whitesides. He started as an assistant professor of chemistry at the University of Washington (Seattle) in 1997 and joined the department of biomedical engineering at Washington University in St. Louis in 2007 as the James M. McKelvey Professor. Since 2012, he holds the position of Brock Family Chair and GRA Eminent Scholar in Nanomedicine at Georgia Tech.

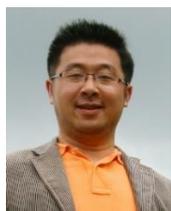

**Hui Zhang** received his Ph.D. in Materials Physics and Chemistry from Zhejiang University in 2005 under the supervision of Professor Deren Yang. He spent two years in the Xia group as a visiting scholar from 2009–2011. He has been a full Professor at Zhejiang University since 2014. His current research interests include synthesis and application of noble-metal nanostructures.

**Table of contents**

**Intermetallic nanocrystals** are superior catalysts over their alloy counterparts with the same composition but a disordered atomic structure. This review article provides an account of recent progress in the development of intermetallic nanocrystals with enhanced catalytic properties for various applications.

Yucong Yan, Jingshan S. Du, Kyle D. Gilroy, Deren Yang, Younan Xia,* and Hui Zhang*

# Intermetallic Nanocrystals: Syntheses and Catalytic Applications

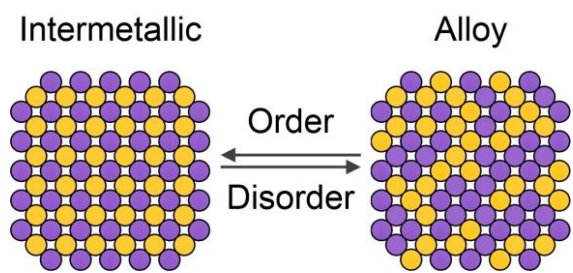